\documentclass[prd,preprintnumbers,amsmath,amssymb]{revtex4}

\usepackage{graphicx}
\usepackage{graphics}
\usepackage{epsfig}
\usepackage{psfrag}
\usepackage{rotating}
\usepackage{dcolumn}
\usepackage{bm}

\newcommand{\beq}{\begin{eqnarray}} 
\newcommand{\eeq}{\end{eqnarray}}

\begin{document}

\preprint{LPT-ORSAY-08-68}

\title{The precision electroweak data in warped extra--dimension models}
\author{Charles Bouchart, Gr\'egory Moreau}
\affiliation{Laboratoire de Physique Th\'eorique, U. Paris--Sud and CNRS, 
F--91405 Orsay, France.}

\begin{abstract}   
The Randall--Sundrum scenario with Standard Model fields in the bulk and a
custodial symmetry is considered. We determine the several minimal quark 
representations allowing to address the anomalies in the forward--backward $b$--quark asymmetry $A^b_{FB}$, while reproducing the bottom and top masses via 
wave function overlaps. The calculated corrections of the $Z \bar b b$
coupling include the combined effects of mixings with both Kaluza--Klein excitations of gauge 
bosons and new $b'$--like states. It is shown that the mechanism, in which the left--handed doublet of third generation quarks results from a mixing on the 
UV boundary of introduced fields $Q_{1L}$ and $Q_{2L}$, is necessary for phenomenological reasons. 
Within the obtained models, both the global fit of $R_b$ with $A^b_{FB}$ [at the various center of mass energies]
and the fit of last precision electroweak data in the light fermion sector can simultaneously be improved significantly with respect to the pure Standard Model case, 
for $M_{KK}=3,4,5$ TeV (first KK gauge boson) and a best--fit Higgs mass $m_h \geq 115$ GeV i.e. compatible with the LEP2 direct limit. 
The quantitative analysis of the oblique parameters $S$,$T$,$U$ 
even shows that heavy Higgs mass values up to $\sim 500$ GeV may still give rise to an acceptable quality of the electroweak data fit, in contrast with the Standard Model.
The set of obtained constraints on the parameter space, derived partly from precision electroweak data, is complementary of a future direct exploration of this 
parameter space at the LHC. 
In particular, we find that custodians, like $b'$ modes, can be as light as $\sim 1200$ GeV i.e. a mass lying possibly in the potential reach of LHC.
\end{abstract}   
\maketitle

\large

\section{Introduction}
\label{intro}

The fine--tuning problem, related to the high discrepancy between the ElectroWeak (EW) symmetry breaking scale
and the Planck mass scale (gauge hierarchy), is probably the strongest indication for the existence of a 
physics underlying the Standard Model (SM). The SM extension to a geome\-trical
setup with additional warped spatial dimension(s), as proposed by Randall and Sundrum (RS) \cite{RS},  
represents a new paradigm 
\footnote{A strongly coupled gauge theory, which remains strongly coupled
in the UV (approaching a non--trivial conformal fixed point) and predicts the Higgs 
as a composite state \cite{MHCM}, has possibly an equivalent description - in the limit of a
large number of colors - in terms of a weakly coupled 5--dimensional theory defined on the truncated 
anti de--Sitter space [AdS/CFT correspondence].}
\cite{AdSCFT} allowing to avoid the fine--tuning problem without supersymmetry. In such an 
higher--dimensional framework, the hierarchy between EW and Planck scales is 
generated exponentially through the warping.

Within the version of the RS model suggested originally - RS1 - the SM fields were confined on the 
TeV--brane where the effective cut--off is of order of the TeV. This feature was entering in conflict
with the presence of dangerous higher dimension operators, inducing Flavor--Changing Neutral Current (FCNC) 
effects, which must be suppressed by energy scales of at least $\sim 10^3$ TeV. Even if
these operators could be reduced by some geometrical factors, the flavor sector was remaining sensitive
to the UltraViolet (UV) physics. It was proposed later \cite{RSloc} to let all SM fields, except the Higgs boson
(then the EW scale remains protected by the low cut--off), propagating along the warped extra dimension.

Furthermore, those RS versions with matter in the bulk benefit from several attractive aspects.
First, a purely geometrical mechanism for generating the fermion flavor structure arises quite naturally: 
if the three families of fermions are localized differently along the extra dimension 
\footnote{As described in Section \ref{model}, the various fermion wave functions along the extra dimension 
can be controlled by 5--dimensional soliton--like masses which are parameterized by dimensionless parameters noted $c_i$,
the index $i$ labeling the fermions.}, their couplings to the Higgs field, and thus 
their 4--dimensional effective Yukawa couplings, exhibit the necessary hierarchical patterns \cite{RSloc,RSmass,RSnu}.
In addition, these RS versions turn out to constitute a suitable framework with respect to model building in general as well as
various specific phenomenological issues. For instance, those allow 
for the unification of gauge couplings at high--energies \cite{UNI-RS}. They even provide new Weakly 
Interacting Massive Particle (WIMP) candidates for the dark matter of universe \cite{LZP2,LZPsym}.

Nevertheless, the fact that the SM fields are located in the bulk leads to the presence of towers
of Kaluza--Klein (KK) excitations associated to fermion and gauge fields. This results in mixings 
between the SM fields (both fermions and bosons) and their KK states. That mixing in turn 
induces tree--level corrections to the SM couplings and hence large deviations to the set of EW observables 
which are measured with an high accuracy nowadays \cite{PDG}. 
Therefore, these measurements impose typically the mass of the first KK gauge boson
excitation $M_{KK}$ (more precisely the KK photon mass) to be larger than 
$\sim 10$ TeV \cite{Burdman}. This bound introduces the little hierarchy problem,
namely the fine--tuning required to explain the smallness of the EW scale with respect to 
$M_{KK}$. However, extending the SM group, by 
gauging the custodial symmetry ${\rm SU(2)_L\! \times\! SU(2)_R\! \times\! U(1)_X}$ 
in the bulk, allows the EW bound on $M_{KK}$ 
to be lowered down to $\sim 3$ TeV \cite{ADMS} \footnote{The same idea was applied to the 
Higgsless models \cite{HiggslessCUSTO}.}. 
An alternative possibility to soften this indirect EW limit on $M_{KK}$ from $\sim 10$ TeV down to $\sim 5$ TeV \cite{EWB}, 
being not retained here, is to consider a scenario using (large) brane--localized kinetic terms for fermions and gauge fields \cite{BraneFB}.

In this paper, we will first show precisely how one can even solve the notorious anomaly on the forward--backward
$b$--quark asymmetry $A^b_{FB}$ in $e^+e^-$ collisions \cite{AFB-SM}, $A^b_{FB}$ constituting with $R_b \equiv \Gamma (Z^0 \to b\bar{b}) / \Gamma (Z^0 \to {\rm hadrons})$
(ratio of the partial decay widths for the $Z^0$ boson) 
the main precision EW observables in the third quark generation sector. Indeed, the $A^b_{FB}$ measurements, around the $Z^0$ pole (LEP1) \cite{PDG,LEPex}, 
at center of mass energies below (from PEP to TRISTAN) \cite{AFB-below,Klaus,AFB-mix} and far above (LEP2) \cite{AFB-above},
remain today the only set of experimental data presenting several significant deviations from the theoretical SM 
predictions. In the RS context with bulk matter, in contrast with supersymmetric models \cite{AFB-SUSY}, tree--level corrections to the 
$Z^0$ boson coupling arise and can be sufficiently large to explain the $A^b_{FB}$ deviations from their SM expectation. Moreover, in the RS model, 
the heavy flavor fermions are localized closely to the TeV--brane
and acquire thus, relatively to light flavors, some larger couplings (mixings) to the $Z^0$ boson (fermion) KK excitations 
which are also typically located towards this same brane.
In turn, the heavier flavor fermions get larger corrections to the $Z^0$ vertex through both boson and fermion mixing effects with KK modes.
Therefore the RS framework can naturally generates deviations of the $Z^0$ couplings, from their SM value, that arise mainly in the third
quark generation sector. Then the Left and right--handed $Z^0$ coupling deviations can be additive in $A^b_{FB}$ while compensating each other 
in the $R_b$ observable, thanks to the sign difference in front of $Z^0$ charges \cite{RSAFB}.

In contrast with the preliminary study in \cite{RSAFB}, here the contribution
to the $Z^0$ coupling correction coming from the mixing between the $b$--quark and the relevant fermionic KK excitations,
a mixing caused by the EW Symmetry Breaking (SB), will be taken into account.
This mixing effect is in general expected to be important since certain fermionic KK excitations are particularly light.
Those are the new right--handed quarks noted $b'_R$ (the so--called `custodians') of electric charge $-1/3$, which correspond 
to some ${\rm SU(2)_R}$ partner of the right--handed top quark $t^c_R$
\footnote{We introduce charge conjugated fields (indicated by the superscript $^c$) in order to define only left--handed SM fermions.}. 
The $b'_R$ quarks have Dirichlet boundary condition on the Planck--brane and Neumann one on the
TeV--brane \cite{Csaki}, which is written $(-+)$, so that they have no zero--mode (the mode with vanishing KK mass).
Their first KK mass tends to be relatively low with respect to the EWSB scale (a few hundred GeV can be reached {\it a priori}) 
since it is controlled \cite{LZP2} by the same $c_{t_R}$ parameter as the top quark which must be
sufficiently small in order to localize $t^c_R$ towards the TeV--brane and thus create a large top mass $m_t$.

Furthermore, we will determine here the list of explicit quark representations under the extended symmetry 
${\rm SU(2)_L\! \times\! SU(2)_R\! \times\! U(1)_X}$, considering only the minimal
representations (not bigger than the representation ${\bf 3}$), that lead to the specific $Z^0$ coupling deviations  
allowing to solve the $A^b_{FB}$ anomaly.
For the various scenarios of representation type obtained, the domains of
parameter space $\{c_i\}$ corresponding to 
solutions of the $A^b_{FB}$ anomaly
and simultaneously reproducing the quark mass values will be presented, a study which was also not performed in \cite{RSAFB}. 
More precisely, we will first show that there exist no values of the parameters
$c_{t_R}$, $c_{b_R}$ and 
$c_{Q_L}$ ($Q_L \equiv (t_L,b_L)^t$ being
the SM ${\rm SU(2)_L}$ doublet field) that reproduce the top and bottom quark masses while addressing the $A^b_{FB}$ problem. 
In order to avoid this difficulty, one is forced to apply a certain mechanism: 
let us suppose that the Yukawa couplings are of the form ${\cal H} \overline{\{t^c_R\}} \{Q_{1L}\}$ and ${\cal H} \overline{\{Q_{2L}\}} \{b^c_R\}$
where e.g. $\{b^c_R\}$ stands for the whole multiplet representation under the custodial symmetry containing the right--handed $b^c_R$ quark
and ${\cal H}$ is the Higgs boson representation. 
The SM left--handed doublet $Q_L$ can result from a combination of $Q_{1L}$ and $Q_{2L}$ which are mixed on the UV boundary. 
Then, there are more freedom on the parameters $c_{Q_{1L}}$ ($=c_1$) and $c_{Q_{2L}}$ ($=c_2$), fixing respectively $m_t$ and $m_b$, than 
on $c_{Q_{L}}$ that fixes both $m_t$ and $m_b$ in the usual case. We will present some regions of the
parameter space $\{c_1,c_2,c_{t_R},c_{b_R}\}$ which satisfy the conditions to solve the $A^b_{FB}$ problem
and also generate the correct $m_t$, $m_b$ values.
\\ The above mechanism was invoked, in the framework of the custodial symmetry ${\rm O(3)}$ \cite{AFB-Rold} 
\footnote{In Section \ref{subO3}, the $A^b_{FB}$ question will be discussed in
the framework of the ${\rm O(3)}$ symmetry \cite{AFB-Rold}.},
as a possibility to consider different group embeddings for $\{Q_{2L}\}$ and thus for $\{b^c_R\}$ leading to a positive correction
of the $Z^0$ coupling. The mechanism was necessary because $\{Q_{1L}\}$ was chosen to be {\it fixed} to a representation 
${\rm ({\bf 2},{\bf 2})_{2/3}}$ under ${\rm SU(2)_L\! \times\! SU(2)_R\! \times\! U(1)_X}$
due to the ${\rm P_{LR}}$ parity, in contrast with the present context where various embeddings for the $\{Q_{1L}\}$ multiplet
will be considered. As explained above, our motivation for applying this mechanism is of a different nature.
A concrete realization of such a mechanism was already proposed in the other context of composite Higgs models \cite{compMECH}.  
The holographic interpretation of this mechanism is that the elementary field $Q_L$ couples to a strongly coupled
Conformal Field Theory (CFT) sector via two different composite operators ${\cal O}_1$ and ${\cal O}_2$, the first responsible
for generating the top mass, the second for the bottom mass \cite{compMECH}.

Finally, within the present RS framework addressing the $A^b_{FB}$ anomaly in the third quark generation sector,
the fit of the precision EW data in the complementary sector of SM gauge bosons and light fermions
will be analyzed as well, in terms of the parameters $S$, $T$ and $U$ which synthesize the corrections to EW observables
\cite{STU}, an analysis missing in \cite{RSAFB}. Some satisfactory $\chi^2$--analysis results
will be obtained for values of $g_{Z'}$ 
\footnote{$g_{Z'}$ denotes the coupling of the $Z'$ boson resulting from a
superposition of $\widetilde W^3$ associated to ${\rm U(1)_R}$,
issued from the ${\rm SU(2)_R}$ breaking, and $\widetilde B$ associated to
${\rm U(1)_X}$.} 
compatible with the $A^b_{FB}$ solutions, and, for
$M_{KK}$ as low as $3-5$ TeV thanks to the global custodial isospin symmetry in the CFT
Higgs sector.
Note that for certain domains of the obtained parameter space, some predicted $b'$ masses are lower: 
those can reach $\sim 1200$ GeV which opens the possibility of a significant direct production rate at the LHC. 
\\ Another encouraging aspect of this obtained fit analysis is the comparison with the SM case. In the SM context,
fits made to high energy precision EW data in 2006, including the $\sin^2\theta_{{\rm eff}}^{{\rm lept}}$ 
\footnote{$\theta_{{\rm eff}}^{{\rm lept}}$ denotes the {\it Weinberg angle} modified by radiative corrections 
and the exponent ``lept'' means that it is the value which can be obtained directly from lepton asymmetry measurements.}
value derived from the $A^b_{FB}$ measurement at LEP1, led to a best--fit value for the Higgs 
mass of $m_h=85_{{\rm +39}}^{{\rm -28}}$ GeV \cite{LEP2006}. 
By removing from the global SM fit the $\sin^2\theta_{{\rm eff}}^{{\rm lept}}$ value coming from $A^b_{FB}$
\footnote{The exclusion from the global fit, of the $\sin^2\theta_{{\rm eff}}^{{\rm lept}}$ value deduced from $A^b_{FB}$, could be 
justified by assuming that the $A^b_{FB}$ measurement suffers from some systematic errors which would have been underestimated by the 
experiments [note that it is not the philosophy adopted throughout this paper]. 
This assumption would explain the $3.2\sigma$ discrepancy between this $\sin^2\theta_{{\rm eff}}^{{\rm lept}}$ value and the 
one which is extracted from ${\cal A}_\ell (SLD)$ (see the end of Section \ref{conf}).}, 
the obtained best--fit Higgs mass would have been even smaller \cite{AbdelREVI}.
In these fits, the values used for the $W^\pm$ boson and top quark masses were $m_{W^\pm}=80.392 \pm 0.029$ GeV and $m_t=171.4 \pm 2.1$ GeV. 
Updating these masses to the most recent world average values of $m_{W^\pm}=80.398 \pm 0.025$ GeV (combined Tevatron Run II
and LEP2 results) \cite{LEP2007,RunIIMW,LEPex} and $m_t=170.9 \pm 1.8$ GeV (Tevatron Run II data from CDF and D$0$) 
\cite{LEP2007,LEPex,RunIIMT}, the SM best--fit Higgs mass decreases down to $m_h=76_{{\rm +33}}^{{\rm -24}}$ GeV \cite{RunIIMW},
a value in weak agreement with the limit deduced from direct Higgs boson searches at LEP2: $m_h>114.4$ GeV at $95 \% C.L.$ \cite{DirectMH}. 
This disagreement in the SM between the direct lower limit on the Higgs mass and its value most favored by precision EW data - especially 
when the $A^b_{FB}$ measurement is excluded from the fit - may be seen as an 
indication for the existence of a new physics underlying the SM (see e.g. \cite{AFB-mix,Chanowitz}).
In our RS scenario, we find that the best--fit Higgs mass can be higher than the LEP2 limit at $114.4$ GeV,
a result which constitutes a possible way out of the above SM conflict.
\\ Besides, we will show that for $m_h \geq 115$ GeV, the fit of precision EW data is significantly improved with respect to SM due to the
corrections of EW observables from mixings with KK states.
In particular, we find that EW fits better than in the SM can be obtained for $M_{KK}=3-5$ TeV and $m_h$ as high as $190$ GeV, a value in the mass range where
the second dominant Higgs decay becomes the channel $h \to Z^0 Z^0$ which offers possibly a clean and purely leptonic final state signature at LHC. 
Besides, the discovery of such an heavy Higgs would thus constitute an indication in favor of RS--like models and would exclude the Minimal 
Supersymmetric Standard Model in which $m_h \lesssim 140$ GeV [except in the warped 5--dimensional
supersymmetry case \cite{Gautam}], namely the conservative bound (holding even in the large $\tan \beta$ limit) due
to the intrinsic structure of the supersymmetric extended Higgs sector \cite{AbdelREVII}.

The paper is organized as follows. In next section, we remind to the reader the theoretical framework of the RS scenario with matter
in the bulk charged under a custodial symmetry. The corrections to the $Z^0$ coupling induced by mixings with fermion and boson KK excitations
is also given there. In Section \ref{EW}, we study the corrections to EW observables arising in the context of the RS model, through an analysis
in the plan $T$ versus $S$ for a fixed $U$ value. The precision EW data on the heavy quarks, including $A^b_{FB}$, are treated separately in Section 
\ref{Third}. In this part, we also discuss the specific fermion representations and describe precisely the corresponding mass matrices. 
Finally, we conclude in the last section.

\section{RS framework and theoretical tools}
\label{model}

\noindent {\bf Geometrical setup:}
The geometrical setup of the RS model consists of a 5--dimensional theory where the warped extra dimension 
is compactified over a $S^{1}/\mathbb{Z}_{2}$ orbifold. 
As already mentioned, in the version studied here, the SM fields propagate along the
extra spatial dimension, like gravity, whereas the Higgs boson is stuck on the TeV--brane. 
While the gravity scale on the Planck--brane is $M_{\rm Planck}= 2.44\times
10^{18}$ GeV, the effective scale on the TeV--brane $M_{\star}=e^{-\pi
kR_{c}} M_{\rm Planck}$ is suppressed by a warp factor which depends on the curvature
radius of the anti--de Sitter space $1/k$ and on the compactification radius $R_c$.
If the product $k R_{c} \simeq 11$ then $M_{\star}\!=\!{\cal O}(1)$ TeV which allows to
address the gauge hierarchy problem. In the following, we will take $k R_{c} \simeq 10.11$ so that
the maximum value of $M_{KK} \simeq 2.4 k e^{-\pi kR_{c}}$, fixed by the theoretical 
consistency bound $k<0.105 M_{\rm Planck}$, is $10$ TeV in agreement with the
range $M_{KK} = 3-5$ TeV considered in Sections \ref{EW} and \ref{Third}. 
\\ \\
\noindent {\bf 5D fermion masses:}
As usually in this context, a parameter noted $c_i$ is introduced for quantifying the 5--dimensional
solitonic mass, $\pm c_f k$, affected to each fermion multiplet in the fundamental theory. These masses
determine the fermion localizations. For instance, if the parameter $c_i$ decreases,
the associated zero--mode fermion gets a 5--dimensional profile closer to the TeV--brane and acquire
in turn a larger mass after EWSB. It is remarkable that this geometrical mechanism for mass generation 
is possible for absolute values satisfying all $\vert c_i \vert \simeq 1$, i.e. for fundamental mass parameters 
all of the same order as the unique scale of the theory: the reduced Planck mass $M_{\rm Planck} \sim k$.
\\ \\
\noindent {\bf Custodial symmetry breaking:}
As suggested originally in \cite{ADMS},
the SM gauge group is recovered after the breaking of the ${\rm SU(2)_R}$ group into ${\rm U(1)_R}$, by boundary condition
and possibly also by a small breaking of ${\rm SU(2)_R}$ in the bulk as will be discussed in more details
later on. Then the breaking ${\rm U(1)_R \times \rm U(1)_X} \to {\rm U(1)_Y}$ occurs via a Vacuum Expectation Value (VEV) on the UV brane:
the state $\widetilde W^3$, associated to the ${\rm U(1)_R}$ group, mixes with $\widetilde B$, associated to the ${\rm U(1)_X}$ factor, 
to give the SM hypercharge $B$ boson, the orthogonal linear combination being the $Z'$ boson.
The $Z'$ profile mimics a $(-+)$ boundary condition and has thus no zero--mode. Its 
first KK mass is close, in value, to $M_{KK}$.
\\ \\
\noindent {\bf $Z^0$ vertex corrections:}
In the RS framework, the $Z^0$ coupling to fermions receives corrections due to the mixing, caused by EWSB, of the SM $Z^0$ boson with its KK excitations
and with the new $Z'$ boson. We now give the expression for these additive corrections \cite{ADMS} that will be useful for the following. The first term originates 
from the mixing with the $Z^0$ KK excitations noted $Z^{(n)}$ [$n \geq 1$], whereas the second term is due to the mixing with the $Z^{\prime(n)}$ excitations:
\begin{equation} 
\frac{\delta g_{Z^0}^{f_{L/R}}}{g_{Z^0}^{f_{L/R}}} \bigg \vert_{{\rm boson}} \simeq 
\frac{m_{Z^0}^2}{(0.4 M_{KK})^2}
\left[ F(c_{f_{L/R}}) + \frac{1}{4} ( 1-\frac{1}{k \pi R_c}) \right] +
\frac{m_{Z^0}^2}{(0.4 [0.981 M_{KK}])^2}
\frac{g^2_{Z'} Q_{Z'}^{f_{L/R}} Q_{Z'}^h}{g^2_Z Q_{Z^0}^{f_{L/R}} Q_{Z^0}^h} F'(c_{f_{L/R}}).
\label{dgsgBOSON} 
\end{equation}
In this expression, $\delta g_{Z^0}^{f_{L/R}}/g_{Z^0}^{f_{L/R}}$ stands for the relative deviation of the whole SM $Z^0$ coupling to a fermion $f_{L/R}$,
which is associated to a chirality $L/R$ and to a parameter $c_{f_{L/R}}$ (see above). $g_Z=g/\cos \theta_W$, where $\theta_W$ is the {\it Weinberg angle},   
is the SM $Z^0$ coupling constant. In our notations, the SM charge 
\begin{equation} 
Q_{Z^0}^{f_{L/R}} = I_{3L}^{f_{L/R}} - Q^f_{\rm e.m.} \sin^2 \theta_W,  
\end{equation}
where $I_{3L}^{f_{L/R}}$ represents the third component of weak isospin and $Q^f_{\rm e.m.}$ the electric charge, affects the $Z^0$ coupling to $f_{L/R}$.
The new charge
\begin{equation} 
Q_{Z'}^{f_{L/R}} = I_{3R}^{f_{L/R}} - Y^{f_{L/R}} \sin^2 \theta' 
\label{QZprime} 
\end{equation}
is in factor of the $Z'$ coupling to $f_{L/R}$: $I_{3R}^{f_{L/R}}$ is the ${\rm SU(2)_R}$ isospin number, $Y^{f_{L/R}}$ the usual hypercharge
and $\theta'$ the mixing angle between $\widetilde W^3$ and $\widetilde B$. The ${\rm U(1)}$ charges are related through:
\begin{equation} 
Y^{f_{L/R}}= Q_X^{f_{L/R}} + I_{3R}^{f_{L/R}} = Q_{\rm e.m.}^f - I_{3L}^{f_{L/R}}
\label{eq:Ycond} 
\end{equation} 
where $Q_X^{f_{L/R}}$ is the $f_{L/R}$ charge under ${\rm U(1)_X}$. 
For the neutral Higgs boson, $Y= I_{3R} = - I_{3L}$ (see Section \ref{ModelI}) 
so that $Q_{Z'}^h / Q_{Z^0}^h = - \cos^2 \theta'$. Finally, the functions $F(c)$ and $F'(c)$
read as,
$$
F(c)= \frac{1}{1-e^{k \pi R_c (2 c -1)}} \frac{1-2c}{3-2c} 
\bigg[ \frac{5-2c}{4(3-2c)} - \frac{k \pi R_c}{2}  \bigg],
$$
\begin{equation}
F'(c)= \frac{1}{1-e^{k \pi R_c (2 c -1)}} \frac{1-2c}{3-2c} 
\bigg[ - \frac{k \pi R_c}{2}  \bigg].
\label{gen}
\end{equation}
The new mixing angle is given by $\sin \theta' \equiv \tilde g' /g_{Z'}$ with 
$g_{ Z^\prime}^2 = \tilde g^2 + \tilde g^{\prime 2}$, where $\tilde g$ and 
$\tilde g'$ are respectively the ${\rm SU(2)_ R}$ and ${\rm U(1)_{X}}$ 
couplings; the coupling $g'$ of the SM ${\rm U(1)_Y}$ group reads as $g'=\tilde g 
\tilde g' /g_{Z'}$. From these relations, one deduces 
\begin{eqnarray} 
2 \sin^2 \theta' = 1 \pm \sqrt{1- (2 g'/g_{Z'}) ^2},
\ \ \ 
2 \tilde g^2/ g_{Z'}^2=1 \mp \sqrt{1-( 2 g'/ g_{Z'}) ^2}.
\label{thetaP}
\end{eqnarray}

On the other side, the mixing caused by EWSB between a given SM fermion $f_{L/R}$ and some fermionic KK excitation, noted $f^{KK}_{L/R}$, brings a different 
type of correction to this $Z^0 \bar f_{L/R} f_{L/R}$ vertex. The final coupling of the lightest mass eigenstate (identified as the observed particle), 
resulting from the $f_{L/R}-f^{KK}_{L/R}$ mixing, is $\sin^2\theta^{KK} g_Z Q_{Z^0}^{f^{KK}_{L/R}} + \cos^2\theta^{KK} g_Z Q_{Z^0}^{f_{L/R}}$ where 
$\theta^{KK}$ is the associated mixing angle. Note that there is no overlap factor depending on $c_{f_{L/R}}$ or $c_{f^{KK}_{L/R}}$, due to the 
orthonormality condition for (fermion) wave functions and the flat profile of $Z^0$ along the fifth dimension. Therefore, one can write the relative
deviation of the coupling $g_{Z^0}^{f_{L/R}} \equiv g_Z Q_{Z^0}^{f_{L/R}}$ for the SM vertex $Z^0 \bar f_{L/R} f_{L/R}$, induced by such a fermion mixing, as:
$$
\frac{\delta g_{Z^0}^{f_{L/R}}}{g_{Z^0}^{f_{L/R}}} \bigg \vert_{{\rm fermion}} =
\frac{(\sin^2\theta^{KK} g_Z Q_{Z^0}^{f^{KK}_{L/R}} + \cos^2\theta^{KK} g_Z Q_{Z^0}^{f_{L/R}}) - g_Z Q_{Z^0}^{f_{L/R}}}{g_Z Q_{Z^0}^{f_{L/R}}} =
$$
\begin{equation}   
\sin^2\theta^{KK} \frac{Q_{Z^0}^{f^{KK}_{L/R}}-Q_{Z^0}^{f_{L/R}}}{Q_{Z^0}^{f_{L/R}}} =
\sin^2\theta^{KK} \frac{I_{3L}^{f^{KK}_{L/R}}-I_{3L}^{f_{L/R}}}{I_{3L}^{f_{L/R}}-Q^f_{\rm e.m.}\sin^2\theta_W}
\label{dgsgFERMION}   
\end{equation}  
since $f_{L/R}$ and $f^{KK}_{L/R}$ must possess an identical electric charge. 
This description of the fermion mixing effect is quite effective but the calculation approach for fermion mixing angles
will be presented later.
For several fermion KK modes mixing with a given $f_{L/R}$, one has to sum over the different corrections of type (\ref{dgsgFERMION}), 
recovering then the formula (19) of \cite{AFB-Rold}.

\section{Fit of precision EW data}
\label{EW}

\subsection{Corrections on EW observables}
\label{corrections}

The presence of KK excitations of gauge bosons induces a modification, after EWSB, of
the EW gauge boson propagators through vacuum polarization effects. These
modifications are called ``oblique'' corrections (as opposed to ``direct''
vertex and box corrections that modify the form of the interactions themselves)
and are parameterized by the three $S_{\rm RS}$,$T_{\rm RS}$,$U_{\rm RS}$
quantities introduced in \cite{STU}.
The index ${\rm RS}$ here indicates that those are evaluated within the RS model,
and the dimensionless parameters $S_{\rm RS}$,$T_{\rm RS}$,$U_{\rm RS}$ are
defined such that they vanish in the absence of KK gauge boson excitations
(see for instance \cite{TaitRS}).

For the light SM fermions, i.e. excluding the quarks $b$ and $t$, the associated parameters $c_{\rm light}$ are taken larger than $0.5$,
the motivation being to generate small masses \cite{RSmass} and to minimize their couplings to KK gauge boson excitations (and thus corrections 
to EW observables). Then in the low--energy effective Lagrangian, the fermion--Higgs higher--dimensional operators, obtained after having integrated  
out heavy KK modes, get a special form which allows one to redefine their effects into purely oblique corrections \cite{RedefI,RedefII,ADMS}. 
\\In the third quark generation sector, the $c_i$ parameters that we will obtain in our analysis are smaller than 
$0.5$, due typically to the necessity of producing relatively large masses and corrections to the $Z^0$ vertex. By consequence, 
the corrections to EW observables in this sector will be treated separately in Section \ref{Third} through a fit analysis 
independent from the oblique parameters.
\\ Finally, the effects from the effective 4--fermion operators are negligible \cite{ADMS} for $c_{\rm light}>0.5$ and $M_{KK} = 3-5$ TeV 
as we consider throughout this paper.

For $c_{\rm light} > 0.5 + \epsilon$ ($\epsilon \gtrsim 0.1$ suffices) typically, the oblique parameter $S_{\rm RS}$ reads as \cite{ADMS,DavSonPer}
\begin{equation}   
S_{\rm RS} \simeq 2 \pi \bigg ( \frac{2.4 v}{M_{KK}} \bigg ) ^2 
-
k \pi^2 R_c \frac{g^2 + g^{\prime 2} + g_{Z'}^2 \cos^4 \theta'}{16} \bigg ( \frac{2.4 v}{M_{KK}} \bigg ) ^4 
\label{SRS}   
\end{equation}
where $v \simeq 246$ GeV is the Higgs boson VEV. 
The first term in Eq.(\ref{SRS}) is the contribution from fermion--Higgs higher--dimensional operators and it is positive. 
The second term comes from the gauge--Higgs sector and is negative so it tends to decrease $S_{\rm RS}$. However, this term is at order $(v/M_{KK})^4$ 
and is thus smaller than the first one at order $(v/M_{KK})^2$. 
Besides, it is not the only term at order $(v/M_{KK})^4$ but it has the particularity to be enhanced by both the factors $k \pi R_c$ and $g_{Z'}^2$.
\\ the oblique parameter $T_{\rm RS}$ reads as \cite{ADMS}
\begin{equation}   
T_{\rm RS} \simeq k \pi^2 R_c \frac{\tilde g^2}{8 e^2} \frac{\tilde M^2}{k^2} \bigg ( \frac{2.4 v}{M_{KK}} \bigg ) ^2 
\label{TRS}   
\end{equation}
where $\tilde M$ is the $\widetilde W^\pm$ mass originating from the small bulk breaking of ${\rm SU(2)_R}$. 
This expression shows that the bulk custodial symmetry protects the parameter $T_{\rm RS}$ from acquiring large values. 
Another scenario (called Scenario II in \cite{ADMS}) is the case where ${\rm SU(2)_R}$ remains unbroken in the bulk. Then
the dominant contribution to $T_{\rm RS}$ comes from the exchange of $t$ and $b'$ quarks at the one--loop level. 
Here we will not consider this different kind of scenario where $T_{\rm RS}$ is generated radiatively and its estimation
relies on a sum over fermion/boson KK towers depending on the choice of quark representations (see Section \ref{Third}). 
\\ Finally, the parameter $U_{\rm RS}$ can be deduced from the vacuum polarization amplitudes given in \cite{ADMS}
\footnote{The authors thank K.~Agashe for helpful discussions on the derivation of the parameter $U_{\rm RS}$.} :
\begin{equation}   
U_{\rm RS} \simeq k \pi^2 R_c \frac{\tilde g^2}{64} \frac{\tilde M^2}{k^2} \bigg ( \frac{2.4 v}{M_{KK}} \bigg )^4 .
\label{URS}   
\end{equation}
It appears that the bulk custodial symmetry also protects this parameter $U_{\rm RS}$. $U_{\rm RS}$ is non--vanishing
only at the order $(v/M_{KK})^4$ in the KK expansion, in contrast with $S_{\rm RS}$ and $T_{\rm RS}$. Hence, 
for all the values of $g_{Z'}$, $\tilde M$ and $M_{KK}$ that we will consider, the $U_{\rm RS}$ values obtained are 
$\sim 5 \ 10^{-5}$ which is completely negligible compared to $S_{\rm RS}$, $T_{\rm RS}$. In the following analysis,
we will thus fix $U_{\rm RS}$ at zero in an extremely good approximation.

\subsection{Confrontation of the RS model with experimental data}
\label{conf}

Concerning the oblique corrections, if one considers the limit $M_{KK} \gg m_{Z^0}$ 
\footnote{The generic case of a low threshold of new physics compared to the EWSB scale was treated in \cite{Burgess,BPRS}.} 
then all the new effects induced by the RS model on the EW observables can be parameterized
in terms of six real variables. Three of those can be reabsorbed in the definitions of
the input quantities, namely the most accurately measured EW observables: $m_{Z^0}$, the electromagnetic fine--structure
constant $\alpha$ and the Fermi coupling constant $G_\mu$ determined in muon decay. This leaves three independent
variables which can be chosen to be $S_{\rm RS}$, $T_{\rm RS}$ and $U_{\rm RS}$ \footnote{An other possible choice of parameterization \cite{AbdelREVI}
is based on the three variables $\sin^2\theta_{{\rm eff}}^{{\rm lept}}$, $\Delta r_W$ (related to EW gauge boson masses) and $\Delta \rho = \alpha T$ 
\cite{AltarelliI} or equivalently the so--called $\epsilon_{1,2,3}$ \cite{AltarelliII}.}. 
Then the corrections to EW observables measured up to the $m_{Z^0}$ scale can be expressed in function of the three variables $S_{\rm RS}$,$T_{\rm RS}$,$U_{\rm RS}$ only.
For instance, the theoretical expression for the observable $m_{W^\pm}$ reads as
\begin{equation}   
m_{W^\pm} = m_{W^\pm} \vert_{SM} + \frac{\alpha c_0^2}{c_0^2-s_0^2} m_{Z^0}^2 \left ( - \frac{1}{2} \ S_{\rm RS} + c_0^2 \ T_{\rm RS} + \frac{c_0^2-s_0^2}{4 s_0^2} \ U_{\rm RS} \right )  
\label{mWexp}   
\end{equation}
where $m_{W^\pm} \vert_{SM}$ represents the value calculated as accurately as possible within the pure SM. 
$s_0$ ($c_0$) stands for $\sin \theta_0$ ($\cos \theta_0$), $\theta_0$ being the electroweak mixing angle obtained 
in the improved Born approximation, namely by taking into account only the well known QED running of $\alpha$ up 
to the $m_{Z^0}$ scale \cite{LEPex}: 
$$
s_0^2 = \frac{1}{2} \bigg [ 1 - \sqrt{ 1 - 4 \frac{\pi \alpha (m_{Z^0}^2)}{\sqrt{2} G_\mu m_{Z^0}^2} } \bigg ] = 0.23098 \pm 0.00012
$$

\begin{figure}[!ht]
\begin{center}
\includegraphics[width=9.cm]{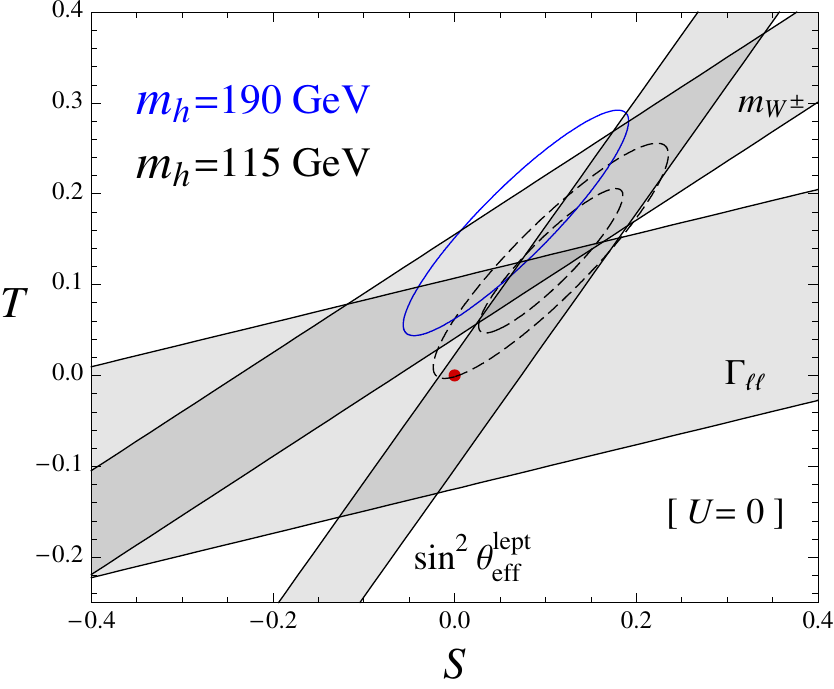} 
\end{center}
\vspace*{-5mm}
\caption{Contour levels in the plan $\{ T,S \}$ at $10.3 \%  $ [$\chi^2/d.o.f. \equiv 1.56$ with a degree of freedom ($d.o.f.$) equal to $11$]
and $14.3 \%  $ [$\chi^2/d.o.f. \equiv 1.45$] for the fit of the 2 experimental values of $m_{W^\pm}$, 
the values of $\Gamma_{ee}$,$\Gamma_{\mu \mu}$,$\Gamma_{\tau \tau}$ and the 6 measurements of $\sin^2\theta_{{\rm eff}}^{{\rm lept}}$:
the two ellipses in dashed--line are for $m_h = 115$ GeV (the external one is at $10.3 \%  $). 
The blue ellipse corresponds to $10.3 \%  $ [$\chi^2/d.o.f. \equiv 1.56$] for $m_h = 190$ GeV.
We also show, for $m_h = 115$ GeV, the three bands corresponding to values of the considered EW observables equal to their measured value (taking into account the
experimental uncertainty) namely: $m_{W^\pm}=80.398 \pm 0.025$ GeV [combined Tevatron Run II and LEP2 results] \cite{LEP2007,RunIIMW,LEPex},
$\Gamma_{\ell \ell}= 83.96 \pm 0.09$ MeV [combined measurements for the 3 flavors] \cite{LEPex} and 
$\sin^2\theta_{{\rm eff}}^{{\rm lept}} = 0.23153 \pm 0.00016$ [from combination of the 6 asymmetries] \cite{LEPex}.
We use the exact results from the two--loop calculations for $m_{W^\pm} \vert_{SM}$ (valid in the domain $100$ GeV $<m_h< 1$ TeV) \cite{MwSM}, 
$\Gamma_{\ell \ell} \vert_{SM}$ ($75$ GeV $<m_h< 350$ GeV) \cite{GllSM} and $\sin^2\theta_{{\rm eff}}^{{\rm lept}} \vert_{SM}$ ($10$ GeV $<m_h< 1$ TeV) \cite{sinSM}. 
The other quantities, on which those SM predictions depend, are fixed at $m_t=170$ GeV (in agreement with recent data: see introduction), 
$\alpha_s=0.118$ \cite{DAtop} and 
$\Delta \alpha=\Delta \alpha_{lept}+\Delta \alpha_{had}+\Delta \alpha_{top}$ with $\Delta \alpha_{lept}=0.03150$ \cite{DAlept}; 
$\Delta \alpha_{had}=0.02758 \pm 0.00035$ \cite{DAhad}; $\Delta \alpha_{top}=-0.00007$ \cite{DAtop}.
Finally, the point [in red] at the origin represents the SM reference point because $U=0$.}
\label{fig:SM}
\end{figure}

\begin{figure}[!ht]
\begin{center}
\includegraphics[width=9.5cm]{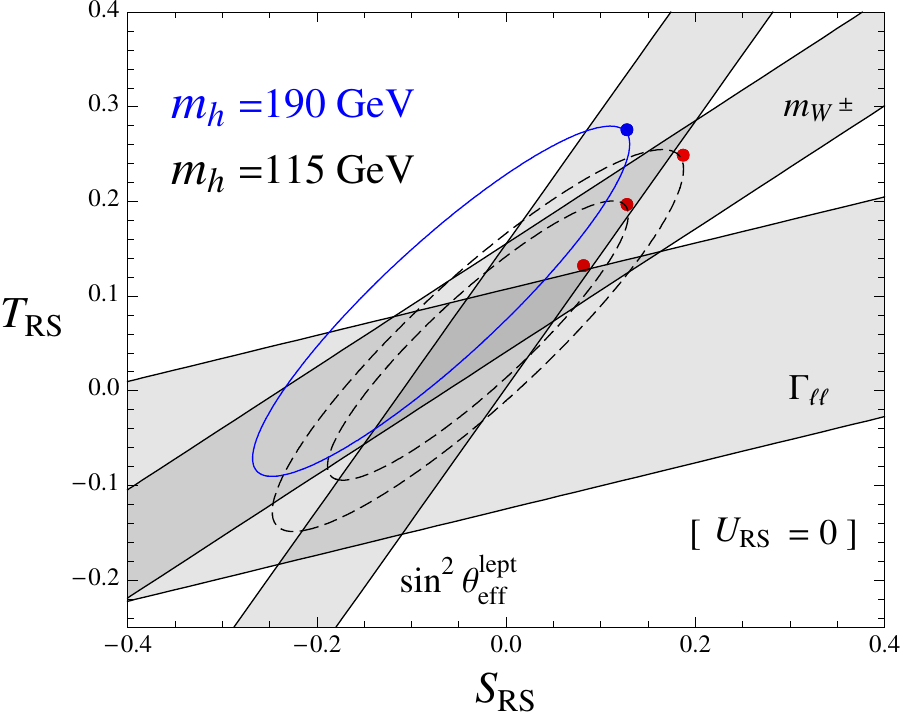}
\end{center}
\vspace*{-5mm}
\caption{Contour levels in the plan $\{ T_{\rm RS},S_{\rm RS} \}$ at $37.3 \%  $ [$\chi^2/d.o.f. \equiv 1.08$ with a degree of freedom ($d.o.f.$) equal to $10$]
and $55.6 \%  $ [$\chi^2/d.o.f. \equiv 0.87$] for the fit of the 2 experimental values of $m_{W^\pm}$, 
the values of $\Gamma_{ee}$,$\Gamma_{\mu \mu}$,$\Gamma_{\tau \tau}$ and the 5 measurements of $\sin^2\theta_{{\rm eff}}^{{\rm lept}}$ (the measurement 
derived directly from the $A^b_{FB}(m_{Z^0})$ experimental value is not included):
the two ellipses in dashed--line are for $m_h = 115$ GeV (the external one is
at $37.3 \%  $).
The blue ellipse corresponds to $42.3 \%  $ [$\chi^2/d.o.f. \equiv 1.02$] for $m_h = 190$ GeV.
We also present, for $m_h = 115$ GeV, the three bands corresponding to values of the considered EW observables equal to their measured value: 
$m_{W^\pm}=80.398 \pm 0.025$ GeV [combined Tevatron Run II and LEP2 results] \cite{LEP2007,RunIIMW,LEPex},
$\Gamma_{\ell \ell}= 83.96 \pm 0.09$ MeV [combined measurements for the 3 flavors] \cite{LEPex} and 
$\sin^2\theta_{{\rm eff}}^{{\rm lept}} = 0.23122 \pm 0.00019$ [from combination of the 5 asymmetries] \cite{LEPex}.
The values used for $m_{W^\pm} \vert_{SM}$, $\Gamma_{\ell \ell} \vert_{SM}$, $\sin^2\theta_{{\rm eff}}^{{\rm lept}} \vert_{SM}$, 
$m_t$, $\alpha_s$ and $\Delta \alpha$ are the same as in Fig.(\ref{fig:SM}).
The three (quasi aligned) theoretical points [in red] correspond respectively, from right to left, to 
$M_{KK}=3$ TeV, $g_{Z'}=2.23$, $\tilde M/k=0.10$ ;
$M_{KK}=4$ TeV, $g_{Z'}=1.25$, $\tilde M/k=0.22$ ;
$M_{KK}=5$ TeV, $g_{Z'}=1.57$, $\tilde M/k=0.18$ [{\it c.f.} Eq.(\ref{SRS})-(\ref{TRS})].
The fourth point [in blue] is for 
$M_{KK}=4$ TeV, $g_{Z'}=1.25$, $\tilde M/k=0.26$. We have set $U_{\rm RS}$ at 
zero (see text).}
\label{fig:RS}
\end{figure}

As is clear e.g. from Eq.(\ref{mWexp}), 
the accurate measurements of EW observables translate into limits in the plan $T_{\rm RS}$ versus $S_{\rm RS}$, if one fixes $U_{\rm RS}$ at zero 
(as justified at the end of previous section).  
These limits depend on the SM expectation, e.g. noted $m_{W^\pm} \vert_{SM}$ in the case of the $W^\pm$ mass. In general, the precise 
predictions of EW observable values
calculated within the SM from QCD/EW corrections depend in turn on the top and Higgs masses as well as the strong coupling constant and
the photon vacuum polarization $\Delta \alpha$ (defined through $\alpha (m_{Z^0}^2)=\alpha (0)/(1-\Delta \alpha)$) \cite{MwSM,GllSM,sinSM}, 
e.g. $m_{W^\pm} \vert_{SM} \equiv m_{W^\pm} \vert_{SM}(m_t,m_h,\alpha_s,\Delta \alpha)$.
\\ \\
\noindent {\bf Numerical results in the SM:}
In Fig.(\ref{fig:SM}), we present the limits in the plan $\{ T,S \}$
\footnote{Here we omit the subscript ${\rm RS}$ in the notation of $S$ and $T$, since the
  experimental limits derived are model--independent (for the considered set of precision EW data)
  and apply in particular for the SM.}
corresponding to values of $m_{W^\pm}$, 
$\sin^2\theta_{{\rm eff}}^{{\rm lept}}$ and the partial $Z^0$ width $\Gamma_{\ell \ell}$ (single charged lepton flavor channel) within
$1 \sigma$ deviation from their experimental central value. The experimental value used here for $\sin^2\theta_{{\rm eff}}^{{\rm lept}}$
is a combination of 6 values resulting from the measurements of the following 6 asymmetries: $A^\ell_{FB}(m_{Z^0})$, ${\cal A}_\ell (P_\tau)$,  
${\cal A}_\ell (SLD)$, $A^b_{FB}(m_{Z^0})$, $A^c_{FB}(m_{Z^0})$ and $Q^{had}_{FB}$. For instance, $A^\ell_{FB}(m_{Z^0}) = (3/4) {\cal A}_e {\cal A}_\ell$ 
is the charged lepton forward--backward asymmetry measured at the $Z^0$ resonance, ${\cal A}_e = (g_{Z^0}^{e_L \ 2}-g_{Z^0}^{e^c_R \ 2})/(g_{Z^0}^{e_L \ 2}+g_{Z^0}^{e^c_R \ 2})$ 
being the pure electron asymmetry parameter \cite{LEPex}.
\\ Note that we concentrate on the experimental measurements of $m_{W^\pm}$, $\sin^2\theta_{{\rm eff}}^{{\rm lept}}$ and $\Gamma_{\ell \ell}$ 
as those are the most precise and crucial in constraining the plan $\{ T,S \}$ \cite{PDG} (these constraints are
clearly model--independent).
\\ In Fig.(\ref{fig:SM}), we also show the contour levels in $\{ T,S \}$ associated to different goodness--of--fit (for $m_h = 115$ GeV and $m_h = 190$ GeV). 
Those result from a $\chi^2$--analysis of the fit between the theoretical predictions for $m_{W^\pm}$, $\sin^2\theta_{{\rm eff}}^{{\rm lept}}$, $\Gamma_{\ell \ell}$ 
(see for example Eq.(\ref{mWexp}))
and their respective experimental values. In this $\chi^2$--analysis, we take into account the 2 measurements of $m_{W^\pm}$ at LEP2 and Tevatron Run II, 
the 3 measurements of $\Gamma_{\ell \ell}$ (one for each flavor) and the 6 measurements of $\sin^2\theta_{{\rm eff}}^{{\rm lept}}$. 
\\ We see on this figure that the reference SM point, at the origin of $\{ T,S \}$ (where e.g. $m_{W^\pm} = m_{W^\pm} \vert_{SM}$),
corresponds to $10.3 \%$ [this probability, used throughout all the paper, represents the p--value quantifying the goodness--of--fit] with respect to the fit of considered
EW observables, for $m_h = 115$ GeV which is close to the direct LEP2 limit \cite{DirectMH}.  
One observes that for $m_h$ increased to $190$ GeV, the fit quality in the SM is degraded. We have also explored the heavy Higgs regime 
and we find that for $m_h=500$ GeV the SM fit has a p--value 
of only $2.5 \ 10^{-9}$ [$\chi^2/11 \equiv 5.7$]. 
\\ \\
\noindent {\bf Numerical results in the RS model:}
In the context of the present RS model, the precision data on $A^b_{FB}$ should be treated separately (see Section \ref{corrections})
so the value of $\sin^2\theta_{{\rm eff}}^{{\rm lept}}$ deduced from the $A^b_{FB}(m_{Z^0})$ measurement, and used in Fig.(\ref{fig:SM}), 
must not be included in the precision EW constraint analysis in the pure gauge and light fermion sector studied in this section. 
More precisely, this value of $\sin^2\theta_{{\rm eff}}^{{\rm lept}}$ equal to $0.23221\pm 0.00029$ is deduced \cite{LEPex} from the experimental value for $A^b_{FB}(m_{Z^0})$
directly by using the SM expression $A^b_{FB}(m_{Z^0}) \vert_{SM} \propto {\cal A}_b = (g_{Z^0}^{b_L \ 2}-g_{Z^0}^{b_R^c \ 2})/(g_{Z^0}^{b_L \ 2}+g_{Z^0}^{b_R^c \ 2})$.
Now in the RS model, $A^b_{FB}(m_{Z^0})$ depends on the corrections induced on $g_{Z^0}^{b_L/R}$ of the type (\ref{dgsgBOSON}) and (\ref{dgsgFERMION}), which are significant
as will be exposed in Section \ref{Third}, so that the $\sin^2\theta_{{\rm eff}}^{{\rm lept}}$ 
value obtained through the SM expression $A^b_{FB}(m_{Z^0}) \vert_{SM}$ is not valid anymore. Therefore, in Fig.(\ref{fig:RS}), we show experimental limits
on $\{ T_{\rm RS},S_{\rm RS} \}$ domains without including the $\sin^2\theta_{{\rm eff}}^{{\rm lept}}$ value extracted through the $A^b_{FB}(m_{Z^0}) \vert_{SM}$ expression 
(the $A^b_{FB}$ measurements will be treated in Section \ref{Third}) 
\footnote{Let us remark that the corrections of type (\ref{dgsgBOSON})-(\ref{dgsgFERMION}), considered for the $b$ quark, apply on the $Z^0$ vertex directly and are thus 
of different form from the oblique contributions, including here the light fermion operator effects.}.
By comparing Fig.(\ref{fig:SM}) and Fig.(\ref{fig:RS}), we observe in particular that this effect of excluding the $A^b_{FB}(m_{Z^0})$ measurement is a shift along the $S$ 
axis of the $1\sigma$ band corresponding to the combined $\sin^2\theta_{{\rm eff}}^{{\rm lept}}$ experimental value (which is modified).

Let us make a comment, at this level, concerning our choice of $m_h$ range in Fig.(\ref{fig:RS}).
We remark that the LEP2 limit $m_h>114.4$ GeV obtained within the SM remains valid in the RS context as a
good approximation. Indeed, it originates from the experimental search for the
Higgs boson production which occurs mainly through the Higgsstrahlung process 
$e^+e^- \to h Z^0$ \cite{DirectMH}. In the RS scenario, the relative deviation of the $h Z^0 Z^0$ 
coupling with respect to the SM is typically of the order of the
percent. For such a small correction, the constraint remains $m_h \gtrsim 115$
GeV (see also \cite{Lillie}) as
shows the Higgs mass limit given as a function of an effective $h Z^0 Z^0$
coupling value, obtained in \cite{DirectMH}.

In Fig.(\ref{fig:RS}), 
we also draw the points corresponding to the theoretical predictions for the values of $S_{\rm RS}$ and $T_{\rm RS}$  
in the RS framework: we see that for $M_{KK}$ as low as $3$ TeV and $g_{Z'}=4\sqrt{\pi / kR_c} \simeq 2.23$ (typical perturbativity limit 
\footnote{See the perturbativity condition for $g_{Z'}$ in \cite{RSAFBLHC} or the expansion parameter in \cite{MHCM}. 
The minimum $g_{Z'}$ value is equal to $2 g' \simeq 0.71$ for consistency reasons (see Eq.(\ref{thetaP})).}), 
the theoretical point in $\{ T_{\rm RS},S_{\rm RS} \}$ still belongs to a
reasonable fit quality region at $37.3 \%  $ [$\chi^2/10 \equiv 1.08$] for $m_h = 115$ GeV (close to
the LEP2 limit).
Taking $g_{Z'}$ at its highest value and a minus sign in the $\sin^2 \theta'$ expression (\ref{thetaP})
allows to optimize the effect of the second (negative) term in Eq.(\ref{SRS}).
This tends to decrease $S_{\rm RS}$ and in turn to improve the fit as exhibits Fig.(\ref{fig:RS}). For the
three other theoretical points of the figure, with larger $M_{KK}$ and smaller $g_{Z'}$, the second term of Eq.(\ref{SRS}) has no 
significant effect. Concerning the $T_{\rm RS}$ parameter, its expression depends on $M_{KK}$ and $g_{Z'}$ (via $\tilde g$)  
but also, in contrast with $S_{\rm RS}$, on the ${\rm SU(2)_R}$ breaking mass $\tilde M$ (see Eq.(\ref{TRS})) 
so that its amount can be controlled independently of $S_{\rm RS}$. 
Here we systematically choose an $\tilde M$ value (and in turn a $T_{\rm RS}$ value) 
which optimizes the EW fit, for a given $S_{\rm RS}$ value. It is remarkable that all the $\tilde M$ values obtained in this way
have an order of magnitude close to $k$, so that no new energy scale is introduced in the RS scenario (which has typically a unique 
fundamental scale), but are smaller than $k$ which guarantees a small breaking of the custodial--isospin in the bulk.
We see on Fig.(\ref{fig:RS}) that for $M_{KK}$ increased up to $4$ TeV, the theoretical predictions for $S_{\rm RS}$ and $T_{\rm RS}$ can
reach the domain at $55.6 \%  $ for $m_h = 115$ GeV, and the EW fit is even
more improved for $M_{KK}=5$ TeV 
due to the smaller $S_{\rm RS}$ 
contribution ({\it c.f.} Eq.(\ref{SRS})). Motivated by the little hierarchy
problem, we do not consider $M_{KK}$ values larger than $5$ TeV.
\\ Now increasing $m_h$ results in a shift of the $\chi^2$ ellipses towards larger $T_{\rm RS}$ and smaller $S_{\rm RS}$ values, so that the fit
would get worse for the three fixed points at $M_{KK}=3,4,5$ TeV discussed
above (and almost aligned in Fig.(\ref{fig:RS})) if $m_h > 115$ GeV.
With $M_{KK}=4$ TeV (which determines $S_{\rm RS}$), while the best--fit point
reachable (by controlling $T_{\rm RS}$) is at $55.6 \%  $ for $m_h = 115$ GeV, 
it is only at $42.3 \%  $ for $m_h = 190$ GeV as shown in
Fig.(\ref{fig:RS}). However, for $m_h$ as large as $190$ GeV, 
the point associated to $M_{KK}=4$ TeV and $\tilde M/k=0.26$ still gives
rise to an acceptable EW fit at $42.3 \%  $ which corresponds to $\chi^2/10 \equiv 1.02$.
The heavy Higgs regime has even been explored: we obtain that the RS fit is at
$25.3 \%  $ [$\chi^2/10 \equiv 1.25$] 
for $m_h=500$ GeV, $M_{KK}=4$ TeV, $g_{Z'}=1.25$ and $\tilde M/k=0.32$. 
\\ \\
\noindent {\bf Comparison SM/RS:}
By comparing Fig.(\ref{fig:SM}) and Fig.(\ref{fig:RS}), we conclude that a
better EW fit can be achieved in the case of the RS scenario than in the SM
situation, for $M_{KK}=3,4,5$ TeV and $m_h = 115$ GeV. This is also true for 
$M_{KK}=4$ TeV and $m_h = 190$ GeV since a fit at $42.3 \%  $ can be reached 
in the RS framework whereas it is worse than $10.3 \%  $ in the SM, as show
the two figures. We have verified that the fit can be improved in the RS
scenario, compared to the SM, for $M_{KK}=3,4,5$ TeV, any possible $g_{Z'}$ value
and $m_h \geq 115$ GeV.
\\ There are two reasons for this improvement. First, there is possibly a positive contribution
to the $T$ parameter in the RS context, in contrast with the SM where $T$ vanishes (like $S$).
This contribution allows to have theoretical predictions for $S_{\rm RS}$, $T_{\rm RS}$ that
reach regions of better goodness--of--fit, since the best--fit point is in a region where
$T$ is positive (see the two figures). The second reason is that, generically, the theoretical
points in the $\{ T_{\rm RS},S_{\rm RS} \}$ plan of Fig.(\ref{fig:RS}) belong to domains
of higher goodness--of--fit, relatively to Fig.(\ref{fig:SM}) [at comparable distances from
the absolute best--fit point]. Indeed, the best--fit point at the center of ellipses in 
Fig.(\ref{fig:RS}) corresponds to a level of agreement of $78.1 \%$ ($77.2 \%$) whereas it
is only at $17 \%$ 
($16.5 \%$) in Fig.(\ref{fig:SM}) for $m_h = 115$ GeV ($190$ GeV).
This is explained by the fact that the fit results, obtained for the SM case in Fig.(\ref{fig:SM}), 
are degraded due to the inclusion in the EW data set of the $\sin^2\theta_{{\rm eff}}^{{\rm lept}}$ value  
deduced {\it directly} from $A^b_{FB}(m_{Z^0})$ by using its SM definition (see above); this $\sin^2\theta_{{\rm eff}}^{{\rm lept}}$ 
value at $0.23221\pm 0.00029$ differs by more than three standard deviations from the other experimental value $0.23098\pm 0.00026$
which is extracted via the ${\cal A}_\ell (SLD)$ measurement (these two $\sin^2\theta_{{\rm eff}}^{{\rm lept}}$ values
are the most precise among the six individual measurements) \cite{LEPex}.
\\ \\
\noindent {\bf Best--fit $m_h$:}
Another important aspect is that 
there exist values of the parameters $M_{KK}$, $g_{Z'}$ and $\tilde M/k$ for which the best EW fit
occurs for an Higgs mass in agreement with its direct LEP2 limit.
For instance, with the values corresponding to the blue point in the plan $\{ T_{\rm RS},S_{\rm RS} \}$ 
of Fig.(\ref{fig:RS}), the best--fit value of $m_h$ is typically $\sim 190$ GeV 
\footnote{Our goal here is to discuss the main variations of the best--fit $m_h$ in RS, and so we do
not compute precisely the best--fit $m_h$ value (including estimated theoretical errors from unknown
higher--order corrections) which is parameter--dependent.}
as can be seen from the oblique shift of ellipses due to an $m_h$ variation (see also e.g. \cite{PDG}).
Indeed, moving away from $m_h = 190$ GeV, the blue ellipse at $42.3 \%  $ would be translated such 
that the fixed theoretical blue point would become located outside this ellipse and would then 
correspond to a quality of the fit of precision EW data lower than $42.3 \%  $.
\\ Therefore, within our RS scenario, the best--fit value of $m_h$ [the $\sin^2\theta_{{\rm eff}}^{{\rm lept}}$
value, derived {\it directly} from $A^b_{FB}(m_{Z^0})$, being not included in this fit] can be higher than the LEP2 
limit of $114.4$ GeV, in contrast with the pure SM context discussed in the introduction.
The reason for this increase of the best--fit Higgs mass with respect to the SM is that the
parameter $T_{\rm RS}$ gets a positive contribution due to the bulk breaking of the custodial 
symmetry, which thus appears to be important here. Indeed, the $S$, $T$ values for the SM  
vanish, and correspond to the origin of Fig.(\ref{fig:RS}), so that the best EW fit is clearly 
reached for $m_h$ below $115$ GeV as recalled in Section \ref{intro}.

\section{The third quark generation sector}
\label{Third}

\noindent {\bf Solving the $A^b_{FB}$ anomaly:}
In the sector of third generation quarks, the most crucial EW constraints come
from the precise measurements of the EW observables $A^b_{FB}$ and $R_b$.  
The deviation, between the measured value of $A^b_{FB}(m_{Z^0})$ at the $Z^0$
pole [$A_{FB}^b=0.0992 \pm 0.0016$] and its SM expectation 
[$A_{FB}^b=0.1037 \pm 0.0008$ for a reference value of the Higgs boson mass at $m_h=129$ GeV] 
\cite{LEPex}, reaches nearly the 3$\sigma$ level and 
is thus the highest one among all EW data. 
Now if some KK excitations induce large corrections to the $Z^0$ vertex of order 
$(\delta g_{Z^0}^{b_R^c}/g_{Z^0}^{b_R^c}) \sim +30 \%$ and 
$(\delta g_{Z^0}^{b_L}/g_{Z^0}^{b_L}) \sim -1 \%$ \cite{RSAFB}, 
then this $A^b_{FB}(m_{Z^0})$ anomaly is addressed, by reducing the 
theoretical asymmetry prediction, while keeping the $R_b$ prediction ($R_b=0.2158$ in
the SM) in agreement with the experimental value: $R_b = 0.21629 \pm 0.00066$ \cite{LEPex}.
It was also shown in \cite{RSAFB} that for such vertex corrections, the global
fit, of the data on $R_b$ and all the $A^b_{FB}$ measurements at various center
of mass energies, is significantly improved as well with respect to the SM case.
In this part of the paper, we are going to show that these amounts of vertex 
corrections can effectively be induced through the two simultaneous mixing
effects arising in the RS model: the mixing of the $b$--quark with fermionic 
KK excitations [of type (\ref{dgsgFERMION})] and the mixing between
$Z^0$ and $Z^{(n)}$,$Z^{\prime (n)}$ [see Eq.(\ref{dgsgBOSON})].

First, we point out a problem of incompatibility of the $c_i$ parameter
values arising when one tempts simultaneously to reproduce the correct $b,t$ masses 
(via the wave function overlap mechanism) 
and the wanted shifts $\delta g_{Z^0}^{b_{L/R}}/g_{Z^0}^{b_{L/R}}$ described
just above. The SM $t^c_R$ quark is generically a single component of a larger multiplet (noted
$\{t^c_R\}$) embedded in a certain representation under the enhanced EW gauge group
${\rm SU(2)_L\! \times\! SU(2)_R\! \times\! U(1)_X}$. 
The $t^c_R$ partners
\footnote{We do not consider the embedding $\{t^c_R\} \equiv ({\bf 1},{\bf 1})_{2/3}$ (option without
any $t^c_R$ partner) which would force one to
take $\{Q_L\} \equiv ({\bf 2},{\bf 2})_{2/3}$ and in turn $b^c_R \in ({\bf 1},{\bf 3})_{2/3}$ to have
gauge invariant Yukawa couplings. In such a configuration, $I_{3R}^{b^c_R}=-1$ leading systematically to either 
a negative or negligible $\delta g_{Z^0}^{b^c_R} / g_{Z^0}^{b^c_R} \vert_{{\rm boson}}$. The problem then is
that $\delta g_{Z^0}^{b^c_R} / g_{Z^0}^{b^c_R} \vert_{{\rm fermion}}$ cannot be large enough to create a
compensation and allow $\delta g_{Z^0}^{b^c_R} / g_{Z^0}^{b^c_R} \vert_{{\rm TOTAL}} \sim +30 \%$, 
as required to address the $A^b_{FB}$ anomaly.}, 
belonging to the multiplet $\{t^c_R\}$, have $(-+)$ boundary
conditions and do not possess zero--modes. The first KK excitation of these
custodians, like e.g. a $b'_R$, has a mass decreasing with $c_{t_R}$ \cite{LZP2}, 
the 5--dimensional parameter of the multiplet $\{t^c_R\}$. The search at colliders
for exotic quarks induces a lower bound on this mass typically around $\sim 300$ GeV
\cite{PDG} which translates into $c_{t_R} \gtrsim -0.5$ [for $M_{KK}=3$ TeV]. 
With such a $c_{t_R}$ range allowed, one is forced to consider $c_{Q_L}
\lesssim 0.25$ in order to generate a correct top quark mass [the 5-dimensional 
Yukawa coupling constants are taken real for simplicity, and, universally equal to $k^{-1}$ here 
to avoid the introduction of new energy scales in the RS theory]. The limit
$c_{Q_L} \lesssim 0.25$, in turn, leads to $c_{b_R} \gtrsim 0.60$ if the 
bottom quark mass is to be generated via the same geometrical mechanism.
Now, even with $c_{b_R} \gtrsim 0.55$, the shift 
$(\delta g_{Z^0}^{b_R^c}/g_{Z^0}^{b_R^c}) \vert_{{\rm boson}}$ 
due to $Z^{(n)}$,$Z^{\prime(n)}$ mixings
cannot exceed $\sim +5 \%$ (see Eq.(\ref{dgsgBOSON})) for $M_{KK}=3$ TeV, and,
the reasonable condition $\vert I_{3R}^{b_R^c} \vert \leq 6$ which limits the field content 
of the model. Indeed, choosing a large $c_{b_R}$ tends to localize $b^c_R$ towards
the UV boundary and thus to reduce its 4--dimensional coupling with $Z^{(n)}$,$Z^{\prime (n)}$.
The shift $(\delta g_{Z^0}^{b_R^c}/g_{Z^0}^{b_R^c}) \vert_{{\rm fermion}}$, caused
by the $b$ mixing with $b'$ states (see Eq.(\ref{dgsgFERMION})), cannot increase significantly the global 
$g_{Z^0}^{b_R^c}$ shift since it never exceeds the order of the percent
typically (see later).
\\ \\
\noindent {\bf A possible mechanism:}
A way out of this problem is provided by the following mechanism. We assume that the Yukawa
couplings have of a form in the Lagrangian: ${\cal H} \overline{\{t^c_R\}} \{Q_{1L}\}$ 
and ${\cal H} \overline{\{Q_{2L}\}} \{b^c_R\}$ (where $Q_{iL} \equiv (t_{iL},b_{iL})^t$
[$i=1,2$] are two different ${\rm SU(2)_L}$ doublets of left--handed fields) which must be invariant under the bulk gauge symmetry
${\rm SU(3)_c\! \times\! SU(2)_L\! \times\! SU(2)_R\! \times\! U(1)_X}$.
Hence, the top mass is controlled by the parameters $c_1$ and $c_{t_R}$ whereas the bottom mass
is determined by $c_2$ and $c_{b_R}$. In other words, the top and bottom masses are fixed independently
via different sets of $c_i$ parameters so that the above problem is avoided. 
\\ The left--handed quarks $t_L$ and $b_L$ must belong to the same SM ${\rm SU(2)_L}$ doublet $Q_L$.
By consequence, $Q_L$ has to be an admixture of $Q_{1L}$ and $Q_{2L}$, which means that there should be
a certain mixing between the whole multiplets $\{Q_{1L}\}$ and $\{Q_{2L}\}$. Now there is no particular reason 
why $\{Q_{1L}\}$ and $\{Q_{2L}\}$ should be embedded into an identical representation and have same quantum numbers 
under ${\rm SU(2)_R\! \times\! U(1)_X}$. 
Therefore, generally, the mixing between $\{Q_{1L}\}$ and $\{Q_{2L}\}$ goes with a breaking of the ${\rm SU(2)_R\! \times\! U(1)_X}$
symmetry through some mass mixing terms. Such mass mixing terms can exist on the UV boundary as ${\rm SU(2)_R\! \times\! U(1)_X}$ is
broken on the Planck--brane (see Introduction and Section \ref{model}). 
\\ A first possibility is that there exist mass terms mixing directly $\{Q_{1L}\}$ and $\{Q_{2L}\}$ on the Planck--brane.
Eitherwise, one can introduce a new field $\{Q_R\}$ localized on the
Planck--brane and having mass terms 
mixing itself with $\{Q_{1L}\}$,$\{Q_{2L}\}$.
Then, while the SM $Q_L$ field would be a combination of the zero--modes of
$Q_{1L}$ and $Q_{2L}$, 
one could get rid of the orthogonal combination
(an extra zero--mode of doublet $\tilde Q_L$) 
e.g. through a mass term mixing it with the field on Planck--brane: $Q_R$. 
\\ \\
\noindent {\bf Fermion representations:}
In the following, we consider scenarios with different representations of $\{Q_{1L}\}$,
$\{Q_{2L}\}$, $\{b^c_R\}$ and $\{t^c_R\}$ under the enhanced EW bulk symmetry 
${\rm SU(2)_L\! \times\! SU(2)_R\! \times\! U(1)_X}$. The choice of these
representations determines the two shifts $\delta g_{Z^0}^{b_{L/R}}/g_{Z^0}^{b_{L/R}} \vert_{{\rm boson}}$ and 
$\delta g_{Z^0}^{b_{L/R}}/g_{Z^0}^{b_{L/R}} \vert_{{\rm fermion}}$, and in
particular, their sign. Indeed, $\delta g_{Z^0}^{b_{L/R}}/g_{Z^0}^{b_{L/R}}
\vert_{{\rm boson}}$ is fixed ({\it c.f.} Eq.(\ref{dgsgBOSON})) 
by $Q_{Z'}^{b_{L/R}}$ and in turn ({\it c.f.} Eq.(\ref{QZprime})) by $I_{3R}^{b_{L/R}}$. 
On the other side, $\delta g_{Z^0}^{b_{L/R}}/g_{Z^0}^{b_{L/R}} \vert_{{\rm fermion}}$
relies ({\it c.f.} Eq.(\ref{dgsgFERMION})) on the isospin
$I_{3L}^{b^{KK}_{L/R}}$ of fermion KK excitations (with
electric charge $-1/3$ like e.g. a $b'_{L/R}$) mixing with $b_{L/R}$.
This shift due to fermion mixing also depends upon the mixing angle $\theta^{KK}$
(introduced in Eq.(\ref{dgsgFERMION})) that results from the mass matrix
for $b_{L/R}$ and KK excitations of identical electric charge. Now this mass
matrix originates from the Yukawa couplings which involve Clebsch--Gordan
coefficients due to products of the chosen top/bottom quark representations. 
\\ More precisely, the first three models that will be considered in next
subsections constitute the obtained exhaustive list of minimal top/bottom
representations (i.e. not larger than the multiplet ${\bf 3}$ under an
individual symmetry)
\footnote{The group structure is minimal in the sense that, throughout this paper, we do not consider
field representations e.g. of the kind $({\bf 1},{\bf 2})_X \oplus ({\bf 3},{\bf 2})_X$ 
[under ${\rm SU(2)_L\! \times\! SU(2)_R\! \times\! U(1)_X}$].} 
which both (1) lead to the above global shift values of 
$\delta g_{Z^0}^{b_{L/R}}/g_{Z^0}^{b_{L/R}}$ allowing to address the
$A^b_{FB}$ anomaly [i.e. the global fit of $A^b_{FB}$ and $R_b$ is required to
be better than $10 \%  $ : see later]  
(2) permit to reproduce good $m_b$ and $m_t$ values. An additional requirement
is that the typical widths of obtained parameter space domains, fulfilling both (1) and (2), 
are not neglectable with respect to the parameters themselves (roughly $\Delta c/|c| = {\cal O}(1)$).
\\ We finish this part by commenting on light fermion representations: one option is to embed 
the other fermion families - namely the first two quark generations plus leptons - into the same ${\rm SU(2)_L\! \times\! SU(2)_R}$ 
representations as $\{Q_{1L}\}$, $\{Q_{2L}\}$, $\{b^c_R\}$ and $\{t^c_R\}$
respectively, thus with a similar duplication mechanism through the
introduction of two left--handed ${\rm SU(2)_L}$ doublets having identical [SM] quantum
numbers (both doublets associated to the $c$--parameter characteristic of light fermions:
$c_{\rm light}>0.5$). Anyway, the choice of these (light) fermion representations does not affect
the study of Section \ref{EW} on the $S_{\rm RS}$,$T_{\rm RS}$,$U_{\rm RS}$ quantities
parameterizing KK gauge boson mixing effects.
Indeed, the $Q_{Z'}^{f_{L/R}}$ charge for SM light fermions $f_{L/R}$, depending on $I_{3R}^{f_{L/R}}$ 
(see Eq.(\ref{QZprime})), does not play any role in computed corrections on EW observables
as the whole $Z'$ coupling to SM light fermions vanishes due to the wave function overlap
factor in the regime of large $c_{\rm light}$ \cite{RSAFB}.

\subsection{Model I}
\label{ModelI}

Within our first model, the third generation quark representations/charges under 
${\rm SU(2)_L\! \times\! SU(2)_R\! \times\! U(1)_X}$ are:
\begin{eqnarray}
\{Q_{1L}\} \ \equiv \ 
({\bf 2},{\bf 1})_{1/6} \ = \
\left(\begin{array}{c} t_{1L} \\ b_{1L} \end{array}\right) 
\ \ \ 
\{t^c_R\} \ \equiv \ 
({\bf 3},{\bf 2})_{1/6} \ = \
\left(\begin{array}{cc} q^{c\prime}_{{\rm (5/3)} R} & t^{c\prime}_R \\ t^c_R & b^{c\prime\prime}_R \\ b^{c\prime}_R & q^{c\prime}_{{\rm (-4/3)} R} \end{array}\right)
\label{rep:Q1LtcR}
\end{eqnarray}
\begin{eqnarray}
\{Q_{2L}\} \ \equiv \ 
({\bf 2},{\bf 3})_{-5/6} \ = \
\left(\begin{array}{ccc} t_{2L} & b'_L & q''_{{\rm (-4/3)} L} \\ b_{2L} & q'_{{\rm (-4/3)} L} & q'_{{\rm (-7/3)} L} \end{array}\right) 
\ \ \ 
\{b^c_R\} \ \equiv \ 
({\bf 1},{\bf 2})_{-5/6} \ = \
\left(\begin{array}{cc} b^c_R & q^{c\prime\prime}_{{\rm (-4/3)} R} \end{array}\right)
\label{rep:Q2LbcR}
\end{eqnarray}
where e.g. $({\bf 3},{\bf 2})_{1/6}$ indicates a triplet under ${\rm SU(2)_L}$ being also a ${\bf 2}$ representation under ${\rm SU(2)_R}$ 
with a charge $Q^{t_R^c}_X=1/6$ with respect to ${\rm U(1)_X}$. Furthermore, 
the multiplet components, with one or several superscripts $^{\prime}$, are fields with boundary conditions of kind $(-+)$ and do not possess zero--modes, whereas the 
other components obey boundary conditions $(++)$ and their zero--modes constitute SM fields. 
For example, $q''_{{\rm (-4/3)} L}$ represents a left--handed exotic quark with electric charge $-4/3$.
Finally, all these components are written inside multiplets accordingly to the increasing $I_{3R}^{f_{L/R}}$ ($I_{3L}^{f_{L/R}}$) isospin number order,
from right to left (bottom to up) with respect to the ${\rm SU(2)_R}$ (${\rm SU(2)_L}$) group.
\\ Besides, with our notations/conventions, the Higgs boson representation under the custodial symmetry is a bidoublet reading as,
\begin{eqnarray}
{\cal H} \ \equiv \ 
({\bf 2},{\bf 2})_{0} \ = \
\left(\begin{array}{cc} H & i \sigma_2 H^\star \end{array}\right) 
\ = \
\left(\begin{array}{cc} h^+ & h^{0\dagger} \\ h^0 & -h^- \end{array}\right) 
\label{rep:H}
\end{eqnarray}
where $H$ represents the usual SM Higgs doublet under ${\rm SU(2)_L}$. 
\\ Therefore, we see easily that the top/bottom Yukawa couplings, written in the fundamental theory in terms of 5--dimensional fields,
constitute well, with this first choice of representations, some operators 
invariant under the total bulk symmetry ${\rm SU(3)_c\! \times\! SU(2)_L\! \times\! SU(2)_R\! \times\! U(1)_X}$:
\begin{equation}   
{\cal S}_{\rm Yuk.}= \int d^5x \sqrt{G} 
\bigg (
\lambda_t^{5D} \ {\cal H} \overline{\{t^c_R\}} \{Q_{1L}\} \ + \ \lambda_b^{5D} \ {\cal H} \overline{\{Q_{2L}\}} \{b^c_R\} \ + \ H.c.
\bigg ),
\label{5Daction}   
\end{equation}
where $G$ is the determinant of the RS metric and $\lambda_{b,t}^{5D}$ are the 5--dimensional Yukawa coupling constants for $b,t$.

At this level, we make a remark which holds also for next models; there is a possible variation on the bottom quark multiplet $\{b^c_R\}$ with
respect to Eq.(\ref{rep:Q2LbcR}), namely,
\begin{eqnarray}
\{b^c_R\} \ \equiv \ 
({\bf 3},{\bf 2})_{-5/6} \ = \
\left(\begin{array}{cc} t^{c\prime\prime}_R & b^{c\prime\prime\prime}_R \\ 
b^c_R & q^{c\prime\prime}_{{\rm (-4/3)} R} \\ 
q^{c\prime\prime\prime}_{{\rm (-4/3)} R} & q^{c\prime}_{{\rm (-7/3)} R} \end{array}\right)
\label{repVAR:Q2LbcR}
\end{eqnarray}
The new $b^{c\prime\prime\prime}_R$ state, introduced here in the field content, has only a Yukawa mass term of type $\bar b^{\prime(n)}_L b^{c\prime\prime\prime(n)}_R$.
In particular, there is neither direct mixing mass term of the form $\bar b^{(0)}_{2L} b^{c\prime\prime\prime(1)}_R$ [due to the 
${\rm SU(2)_L\! \times\! SU(2)_R}$ structure] nor $\bar b^{(0)}_{1L} b^{c\prime\prime\prime(1)}_R$ [as a Yukawa coupling of type ${\cal H} \overline{\{Q_{1L}\}} \{b^c_R\}$ 
would not be gauge invariant] nor $\bar b^{c(0)}_R b^{c\prime\prime\prime(1)}_L$ [since $b^c_R$ and $b^{c\prime\prime\prime}_R$ are in the same multiplet]. 
Therefore, no strong mixing between the SM $b$ and the new $b^{c\prime\prime\prime}$ arises (it is only an indirect mixing through $b^{\prime}_L$). 
The consequences are that there is neither a noteworthy effect on $m_b$
\footnote{The presence of this new $t^{c\prime\prime}_R$ field has also no significant effect on $m_t$.}
nor a significant contribution to $\delta g_{Z^0}^{b_{L/R}}/g_{Z^0}^{b_{L/R}} \vert_{{\rm fermion}}$.
This is the reason why in case of Eq.((\ref{repVAR:Q2LbcR})), we find (positive) numerical results/plots almost identical to the case (\ref{rep:Q2LbcR})
and we will not present those explicitly.
\\ \\
\noindent {\bf Mass matrix:}
First, we describe the mass matrix for the bottom quark and KK excitations arising with the representations (\ref{rep:Q1LtcR})-(\ref{rep:Q2LbcR}), 
a matrix that turns out to be useful for the following. 
\\ We do not consider the mixing with the first two generations of down quarks
since considering the whole mass matrix would require to specify the three flavor model assumed (including all Yukawa coupling
constant values and the choice of $c$--parameters for different left/right--handed quarks) which has to reproduce quark masses,
a task beyond the scope of our work. The fact of not taking into account the mixing with $d$,$s$ does not affect at all the final shift 
$\delta g_{Z^0}^{b_{L/R}}/g_{Z^0}^{b_{L/R}} \vert_{{\rm fermion}}$ 
since $d$,$s$ have of course the same $g_{Z^0}^{b_{L/R}}$ couplings as the $b$ quark.
Besides, neglecting the $b$ mixing with $d$,$s$ should not a priori modify greatly the bottom mass obtained by diagonalizing the whole
mass matrix cause the $V_{CKM}$ mixing angles are small \cite{PDG}. Nevertheless, in order to take into account this approximation effect, we will
introduce an uncertainty range for the bottom mass when trying to reproduce its value theoretically from the geometrical mechanism (this uncertainty 
is also motivated by later considerations on the energy scale dependence of fermion masses). 
\\ Hence, we present the bottom mass matrix ${\cal M}_b$ in the field basis 
$$\Psi_R^t = (b^{c(0)}_R, b^{\prime(1)}_R, b^{c\prime(1)}_R, b^{c\prime\prime(1)}_R, b^{c\prime(2)}_R, b^{c\prime\prime(2)}_R),
\ \ \
\Psi_L^t = (b^{(0)}_L, b^{\prime(1)}_L, b^{c\prime(1)}_L, b^{c\prime\prime(1)}_L, b^{c\prime(2)}_L, b^{c\prime\prime(2)}_L)$$ where  
the superscript $^{(n)}$ [$n=0,1,2,\dots$] indicates the KK excitation level. We have checked numerically that these sets of lightest 
fermionic KK states are the most important in the calculation of the smallest ${\cal M}_b$ eigenvalue noted $m_b$ (which should be equal to 
the measured value of the bottom quark mass), $m_{b_2}$ [the second smallest ${\cal M}_b$ eigenvalue] and $\delta g_{Z^0}^{b_{L/R}}/g_{Z^0}^{b_{L/R}} \vert_{{\rm fermion}}$. 
It means that heavier modes of the KK towers tend to decouple and taking them into account does not change significantly
the theoretical predictions for $m_b$, $m_{b_2}$ and $\delta g_{Z^0}^{b_{L/R}}/g_{Z^0}^{b_{L/R}} \vert_{{\rm fermion}}$. 
\\ After EWSB, the bottom--like quarks get Dirac masses through the Yukawa couplings (which must be contracted with respect to group representations).
These mass terms read, in the 4--dimensional effective Lagrangian, as [with $c_\theta \equiv \cos \theta$; $s_\theta \equiv \sin \theta$]:
$$
{\cal L}_{\rm mass}^b \! = \! \bar \Psi_L {\cal M}_b \Psi_R \! + \! H.c. \ \ \ \ \mbox{with,} 
$$
\begin{eqnarray}   
{\cal M}_b =
\left(  
\begin{array}{cccccc} 
\tilde v_b c_\theta f^{(0)\star}_{c_2} f^{(0)}_{c_{b_R}} & 0 & \tilde v_t s_\theta f^{(0)\star}_{c_1} g^{(1)}_{c_{t_R}} & \sqrt{2} \tilde v_t s_\theta f^{(0)\star}_{c_1} g^{(1)}_{c_{t_R}} &
\tilde v_t s_\theta f^{(0)\star}_{c_1} g^{(2)}_{c_{t_R}} & \sqrt{2} \tilde v_t s_\theta f^{(0)\star}_{c_1} g^{(2)}_{c_{t_R}} \\  
(\tilde v_b/\sqrt{2}) g^{(1)\star}_{c_2} f^{(0)}_{c_{b_R}} & m^{\prime(1)}_{c_2} & 0 & 0 & 0 & 0 \\  
0 & 0 & m^{\prime(1)}_{c_{t_R}} & 0 & 0 & 0 \\  
0 & 0 & 0 & m^{\prime(1)}_{c_{t_R}} & 0 & 0 \\
0 & 0 & 0 & 0 & m^{\prime(2)}_{c_{t_R}} & 0 \\  
0 & 0 & 0 & 0 & 0 & m^{\prime(2)}_{c_{t_R}}   
\end{array} 
\right)
\label{MbTOTAL}   
\end{eqnarray}
where $\tilde v_{b,t} = \lambda_{b,t}^{5D} v / \sqrt{2} \pi R_c$, 
$m^{\prime(n)}_{c}$ is the $n$--th KK mass for $(-+)$ fields ($m^{\prime(1)}_{c_{t_R}}$ tends to be small if $c_{t_R}$ is reduced, in order to reproduce $m_t$) 
and $\theta$ is the effective angle of the mixing between $b^{(0)}_{1L}$ and $b^{(0)}_{2L}$ resulting in the $b^{(0)}_L$ field with SM quantum numbers.
In the above mass matrix, $f^{(n)}_{c}/\sqrt{2 \pi R_c}$ and $g^{(n)}_{c}/\sqrt{2 \pi R_c}$ stand for the wave functions of the $n$--th KK mode of a field  
characterized by the $c$--parameter and, respectively, $(++)$ and $(-+)$ boundary conditions; all wave function values are taken at the position of the 
TeV--brane, $x_5\!=\!\pi R_c$ (where is confined the Higgs boson). For instance, 
$$f^{(0)}_{c}(x_5)=e^{(0.5-c)k \vert x_5 \vert }\sqrt{(1-2c)\pi k R_c}/\sqrt{e^{(1-2c)\pi k R_c}-1}.$$
Some zeroes among the matrix elements in Eq.(\ref{MbTOTAL}) are due the fact that the $(+-)$ fields 
$b^{c\prime(n)}_L$, $b^{c\prime\prime(n)}_L$ and $b^{\prime(n)}_R$ have a Dirichlet boundary condition on the TeV--brane and hence do not couple to the Higgs boson.
The presence of zeroes is also partly explained by the fact that there is no mass term mixing $b^{\prime(n)}_{R/L}$ with $b^{c\prime(n)}_{L/R}$ or $b^{c\prime\prime(n)}_{L/R}$, 
since there exist no Yukawa coupling between $\{Q_{2L}\}$ and $\{t^c_R\}$ invariant under ${\rm U(1)_X}$ ($\{Q_{2L}\}$ and $\{t^c_R\}$ have different $Q_X$ charges).

\begin{figure}[!ht]
\begin{center}
\includegraphics[width=7.5cm]{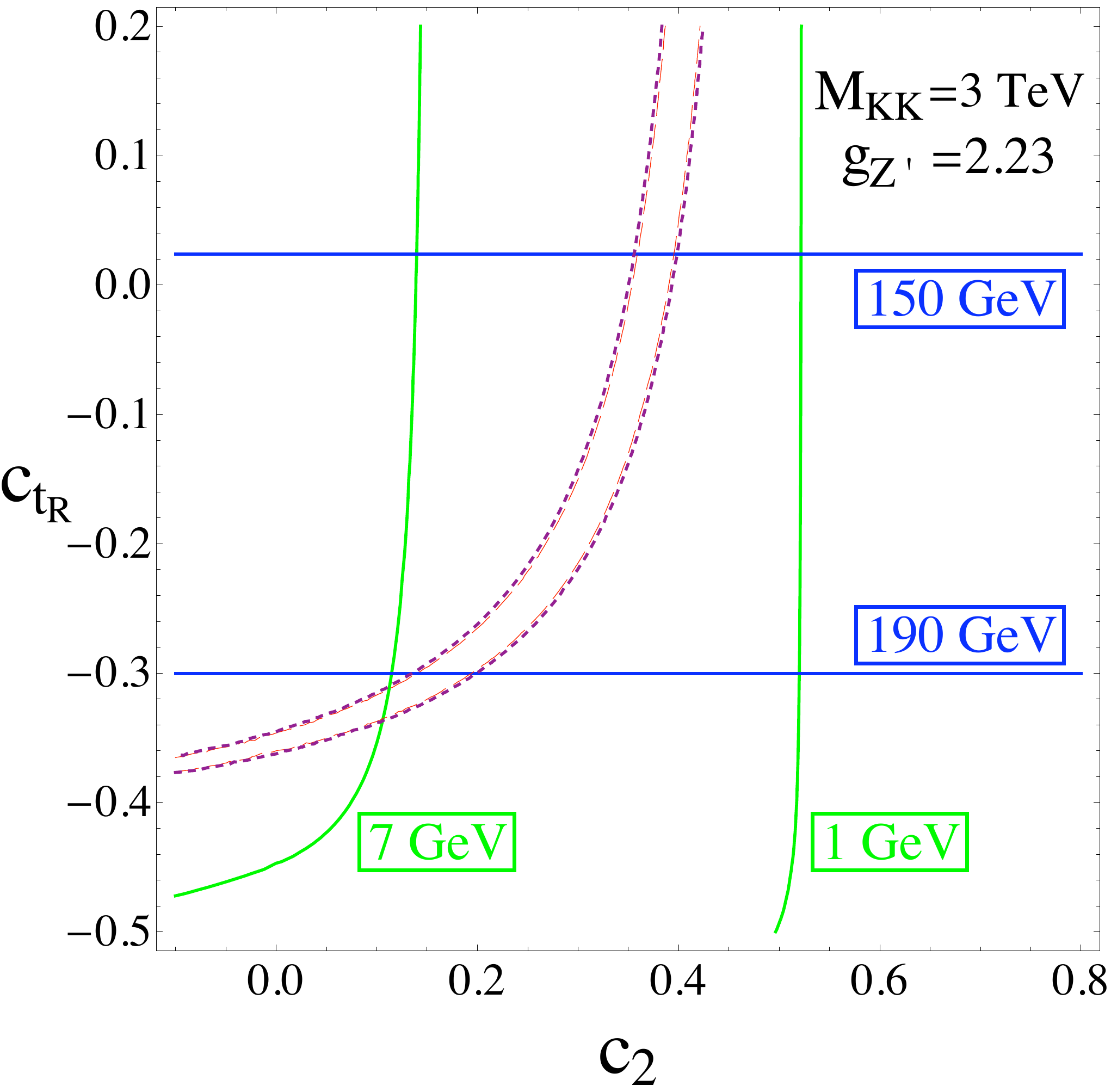} 
\includegraphics[width=7.5cm]{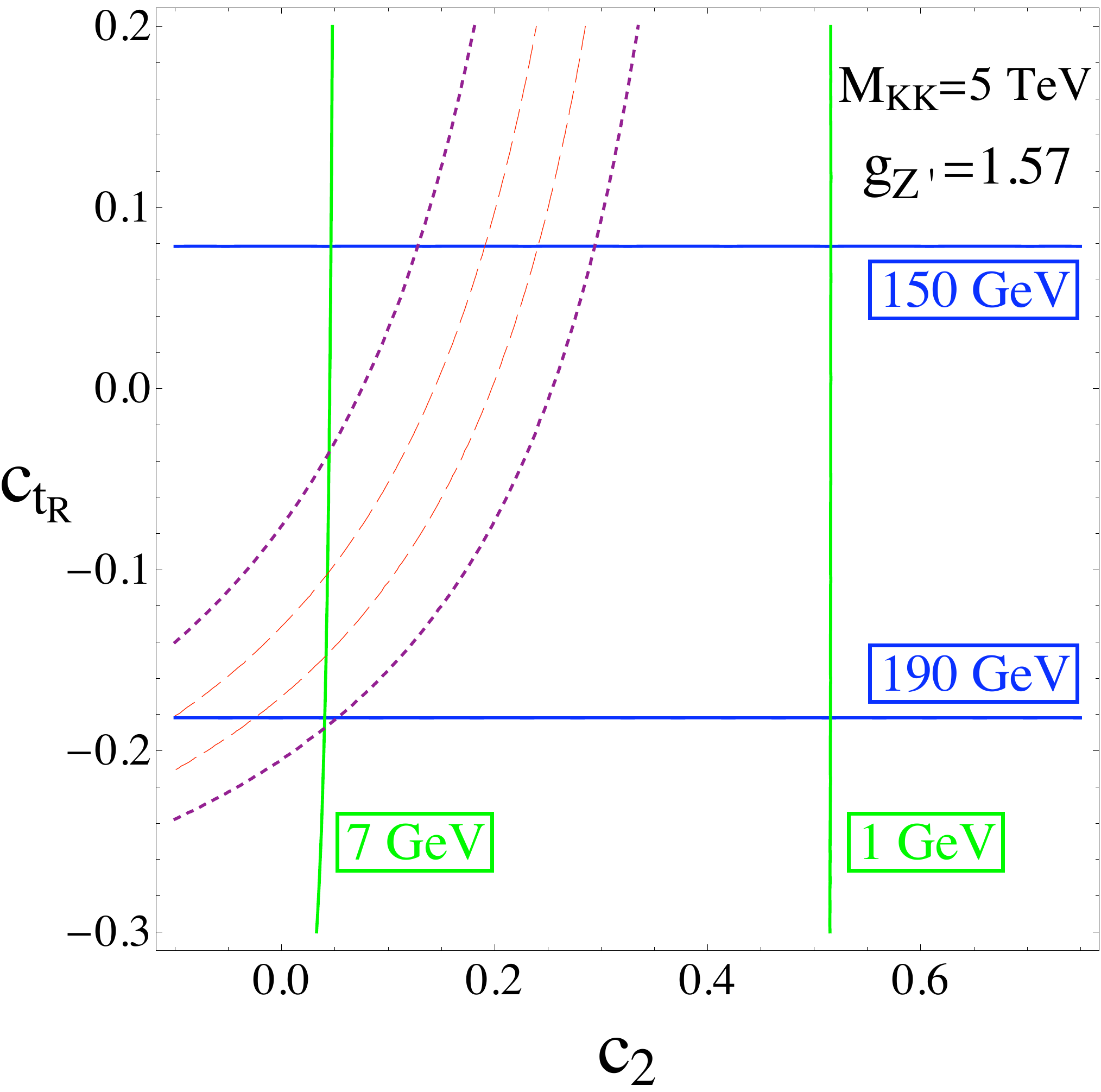} 
\end{center}
\vspace*{-5mm}
\caption{Contour levels in the plan $\{c_{t_R},c_2\}$ at $10 \%  $ [$\chi^2/d.o.f. \equiv 1.63$ with a degree of freedom at $9$] (red dashed--lines)
and $8 \%  $ [$\chi^2/d.o.f. \equiv 1.71$] (purple dotted--lines) for the fit of the experimental data on $R_b$ and the eight measurements of $A^b_{FB}$ at different
center of mass energies. For comparison, in the SM, this fit reaches only $0.8 \%  $ [$\chi^2_{SM}/d.o.f. \equiv 2.47$].
We also present the curves corresponding to $m_b=1;7$ GeV (green plain--lines) and $m_t=150;190$ GeV (blue plain--lines).  
These two plots, associated to different values of $g_{Z'}$ and $M_{KK}$ (as indicated on the figure), are obtained for Model I with 
$\cos^2\theta=0.10$, $\lambda_{t}^{5D} k=3$, $\lambda_{b}^{5D} k=0.4$, $c_{b_R}=0.45$, $c_1=0.47$ [plot for $M_{KK}=3$ TeV];  
$\cos^2\theta=0.10$, $\lambda_{t}^{5D} k=3$, $\lambda_{b}^{5D} k=0.2$, $c_{b_R}=0.28$, $c_1=0.45$ [plot for $M_{KK}=5$ TeV].}
\label{fig:MODI}
\end{figure}

The matrix of Eq.(\ref{MbTOTAL}) is diagonalized by $6 \times 6$ (in a good approximation) unitary matrices $U_{bL/R}$,
via the basis transformation  $\Psi^m_{L/R}\!=\!U_{bL/R}\Psi_{L/R}$ :   
\begin{equation}
U_{bL} {\cal M}_b U_{bR}^\dagger \ = \ {\rm diag}~(m_b,m_{b_2},\dots)
\label{MbEIGEN}  
\end{equation}
$m_b$ having to be associated with the experimental value for the bottom quark mass (the
unitary matrices are chosen such that $m_b\!<\!m_{b_2}\!<\!\dots$). The 
$\Psi^m_{L/R}$ components are the bottom mass eigenstates.  
\\ \\
\noindent {\bf $Z^0$ couplings:}
Now, one is able to derive the $Z^0$ couplings to bottom quarks, first, in the weak basis, and then, in the mass basis. 
In the weak basis, the 4--dimensional effective Lagrangian of the Neutral Current interaction for left--handed bottom quarks is given by,
$$
{\cal L}_{\rm NC}^{b_L} \! = \! Z^\mu_{\rm phys} \bar \Psi_L \gamma_\mu {\cal G}_{bL} \Psi_L  \ \ \ \ \mbox{where,}  \ \ \ \ \ \ \ \ \ \ \ \ \ \ \ \ {\cal G}_{bL} = 
$$
{\small
\begin{eqnarray}   
\left(  
\begin{array}{cccccc} 
{\cal G}_{bL11} & 0 & 0 & 0 & 0 & 0 \\  
0 & g_{Z^0}^{b^{\prime(1)}_L} + \delta g_{Z^0}^{b^{\prime(1)}_L}\vert_{{\rm Z^{\prime (1)}}} & 0 & 0 & 0 & 0 \\  
0 & 0 & g_{Z^0}^{b^{c\prime(1)}_L} + \delta g_{Z^0}^{b^{c\prime(1)}_L}\vert_{{\rm Z^{\prime (1)}}} & 0 & \delta g_{Z^0}^{b^{c\prime(1-2)}_L}\vert_{{\rm Z^{\prime (1)}}} & 0 \\  
0 & 0 & 0 & g_{Z^0}^{b^{c\prime\prime(1)}_L} + \delta g_{Z^0}^{b^{c\prime\prime(1)}_L}\vert_{{\rm Z^{\prime (1)}}} & 0 & \delta g_{Z^0}^{b^{c\prime\prime(1-2)}_L}\vert_{{\rm Z^{\prime (1)}}} \\
0 & 0 & \delta g_{Z^0}^{b^{c\prime(1-2)}_L}\vert_{{\rm Z^{\prime (1)}}} & 0 & g_{Z^0}^{b^{c\prime(2)}_L} + \delta g_{Z^0}^{b^{c\prime(2)}_L}\vert_{{\rm Z^{\prime (1)}}} & 0 \\  
0 & 0 & 0 & \delta g_{Z^0}^{b^{c\prime\prime(1-2)}_L}\vert_{{\rm Z^{\prime (1)}}} & 0 & g_{Z^0}^{b^{c\prime\prime(2)}_L} + \delta g_{Z^0}^{b^{c\prime\prime(2)}_L}\vert_{{\rm Z^{\prime (1)}}}   
\end{array} 
\right)
\label{NCinter}   
\end{eqnarray}
}
$$
\mbox{with} \ \ \
{\cal G}_{bL11} = g_{Z^0}^{b_L} + s_\theta^2 \delta g_{Z^0}^{b_{1L}}\vert_{{\rm boson}} + c_\theta^2 \delta g_{Z^0}^{b_{2L}}\vert_{{\rm boson}} .
$$
Strictly speaking, this Lagrangian describes the interaction of the physical state $Z^0_{\rm phys}$ rather than the SM $Z^0$ boson;
$Z^0_{\rm phys}$ is the lightest eigenstate of the mass matrix mixing $Z^0$,$Z^{(n)}$,$Z^{\prime (n)}$ \cite{LHNC} and its mass is equal 
to the measured mass $m_{Z^0}$.
In Eq.(\ref{NCinter}), e.g. $\delta g_{Z^0}^{b_{1L}}\vert_{{\rm boson}}$ depends on $c_1$ [see Eq.(\ref{dgsgBOSON})] and $Q_{Z'}^{b_{1L}}$ relying on $I_{3R}^{b_{1L}}$
[see Eq.(\ref{QZprime})] which is dictated by the $\{Q_{1L}\}$ representation [see Eq.(\ref{rep:Q1LtcR})]. 
\\For instance, the correction to the $Z^0 \bar b^{c\prime(1)}_L b^{c\prime(1)}_L$ vertex due to gauge boson mixing is different from the correction to the SM vertex $Z^0 \bar b^{(0)}_L b^{(0)}_L$ since
$b^{c\prime(1)}_L$ [$b^{(0)}_L$] is a KK excitation [zero--mode] of a field with $(-+)$ [$(++)$] boundary conditions. 
The $Z^0 \bar b^{c\prime(1)}_L b^{c\prime(1)}_L$ correction, noted $\delta g_{Z^0}^{b^{c\prime(1)}_L}\vert_{{\rm Z^{\prime (1)}}}$ in Eq.(\ref{NCinter}), is at order $(v/M_{KK})^2$
and gives rise in turn (through the fermion mixing generated by ${\cal M}_b$) to a deviation of SM coupling $Z^0 \bar b^{(0)}_L b^{(0)}_L$ at order $(v/m^{\prime(1)}_{c})^2 (v/M_{KK})^2$, which is thus a subleading
effect compared to the direct deviations $\delta g_{Z^0}^{b_{1,2 \ L}}\vert_{{\rm boson}}$ at order $(v/M_{KK})^2$ [{\it c.f.} Eq.(\ref{dgsgBOSON})].
In the calculation of $\delta g_{Z^0}^{b^{c\prime(1)}_L}\vert_{{\rm Z^{\prime (1)}}}$, we have considered the main boson mixing effects, namely the mixings with the first neutral KK states
$Z^{(1)}$ and $Z^{\prime (1)}$, which depend on the free parameters $g_{Z'}$ and $M_{KK}$ \cite{LHNC}.  
$\delta g_{Z^0}^{b^{c\prime(1)}_L}\vert_{{\rm Z^{\prime (1)}}}$ is also determined by the 4--dimensional effective couplings $Z^{(1)} \bar b^{c\prime(1)}_L b^{c\prime(1)}_L$
and $Z^{\prime (1)} \bar b^{c\prime(1)}_L b^{c\prime(1)}_L$ which involve $g_Z$, $g_{Z'}$, $Q_{Z^0}^{b^{c\prime}_L}$, $Q_{Z'}^{b^{c\prime}_L}$ and the $b^{c\prime(1)}_L$ 
location (necessary to compute the wave function overlap between $b^{c\prime(1)}_L$ and $Z^{(1)}$ or $Z^{\prime (1)}$) fixed by the $c_{t_R}$ parameter.
\\ The orthonormality property for fermion wave functions together with the flatness of $Z^0$ profile along the fifth dimension result in a vanishing
overlap factor eliminating e.g. the coupling $Z^0 \bar b^{c\prime(1)}_L b^{c\prime(2)}_L$. However, the coupling $Z^0_{\rm phys} \bar b^{c\prime(1)}_L b^{c\prime(2)}_L$ 
receives a contribution, from the small $Z^{(1)}$ and $Z^{\prime (1)}$ components of the $Z^0_{\rm phys}$ state, which is written as 
$\delta g_{Z^0}^{b^{c\prime(1-2)}_L}\vert_{{\rm Z^{\prime (1)}}}$ in Eq.(\ref{NCinter}).
\\ Moving to the mass basis, the $Z^0_{\rm phys}$ interaction is described by the Lagrangian:
\begin{eqnarray}   
{\cal L}_{\rm NC}^{b_L} \  = \ Z^\mu_{\rm phys} \ \bar \Psi^m_L \ \gamma_\mu \ {\cal G}^m_{bL} \ \Psi^m_L ,  
\ \ \ \mbox{where,} \ \ \
{\cal G}^m_{bL} \ = \ U_{bL} \ {\cal G}_{bL} \ U_{bL}^\dagger .
\label{NCinterMASS}   
\end{eqnarray}
The physical state associated to the bottom quark is the lightest eigenstate of ${\cal M}_b$ with mass $m_b$ and noted: $b_{{\rm phys} L/R}$
(first component of $\Psi^m_{L/R}$). Its coupling to $Z^0_{\rm phys}$ is given by the matrix element ${\cal G}^m_{bL11}$ (first line and first column). 
Hence the relative deviation of the observed coupling $Z^0_{\rm phys} \bar b_{{\rm phys} L} b_{{\rm phys} L}$ with respect to the pure SM coupling $Z^0 \bar b_L b_L$ reads as,
\begin{eqnarray}   
\frac{\delta g_{Z^0}^{b_{L}}}{g_{Z^0}^{b_{L}}} \bigg \vert_{{\rm TOTAL}} 
=
\frac{{\cal G}^m_{bL11}-g_{Z^0}^{b_{L}}}{g_{Z^0}^{b_{L}}}.
\label{dgTOTAL}   
\end{eqnarray}
This shift is a combination of both the effects from neutral gauge boson [type (\ref{dgsgBOSON})] and fermion [type (\ref{dgsgFERMION})] mixings.
\\ A similar analysis of the $Z^0_{\rm phys}$ coupling matrix in the mass basis can be easily performed for the right--handed bottom--like quarks, and, the
shift $\delta g_{Z^0}^{b^c_{R}}/g_{Z^0}^{b^c_{R}} \vert_{{\rm TOTAL}}$ is derived through an analog method.
\\ In the present model, $\delta g_{Z^0}^{b_R^c} / g_{Z^0}^{b_R^c} \vert_{{\rm boson}}$, given by Eq.(\ref{dgsgBOSON}), is positive for relevant $c$ values. 
Besides, the formula (\ref{dgsgFERMION}) indicates that the individual mixings of $b^{\prime(1)}_R$, $b^{c\prime(1,2)}_R$, $b^{c\prime\prime(1,2)}_R$ with $b^{c(0)}_R$
give different sign contributions to the shift $\delta g_{Z^0}^{b_R^c} / g_{Z^0}^{b_R^c} \vert_{{\rm fermion}}$. Adding all these fermion mixings together lead to 
a positive global correction $\delta g_{Z^0}^{b_R^c} / g_{Z^0}^{b_R^c} \vert_{{\rm fermion}}$. Hence, the combined effect of fermion and boson mixings gives rise 
to $\delta g_{Z^0}^{b_{R}^c}/g_{Z^0}^{b_{R}^c} \vert_{{\rm TOTAL}} > 0$, as wanted to address the $A^b_{FB}$ anomaly question.
In contrast, we find here that $\delta g_{Z^0}^{b_L} / g_{Z^0}^{b_L} \vert_{{\rm boson}} <0$ and $\delta g_{Z^0}^{b_L} / g_{Z^0}^{b_L} \vert_{{\rm fermion}} >0$, 
whereas $\delta g_{Z^0}^{b_{L}}/g_{Z^0}^{b_{L}} \vert_{{\rm TOTAL}} < 0$ as required.
\\ \\
\noindent {\bf Flavor structure and the $A^b_{FB}$ anomaly:}
Before presenting numerical results and exploring the parameter space, one needs to discuss a potential problem caused by the bottom quark which occurs
generically in RS models with bulk matter: the dangerous down--quark FC couplings to the KK gluon excitations, induced by the non--universality in the interaction
basis \cite{RSloc,RSmass}, lead to tree--level contributions to mass splittings for the $B$ meson and in the Kaon system which become large for $M_{KK}$ as
low as the TeV scale \cite{RSmass,RSFlav}. In particular, for $M_{KK}=3$ TeV, theoretical predictions for the CP violation effect in the Kaon system, $\epsilon_K$, 
seem to conflict with experimental data \cite{CFW}. 
\\ Let us recall the origin of this problem. In case the three generations of SM down--quarks are localized towards the Planck--brane (i.e. having 
$c_{\rm light}$ {\it and} $c$--parameters for the $b$ larger than $0.5$), they possess quasi--universal couplings to first KK gluon excitations \cite{HRizzoBIS}, 
due to the flat profiles of these KK states near the UV boundary, and thus almost no FC couplings are induced in the mass basis (RS--GIM mechanism). 
However, taking both $c_{Q_L}$ and $c_{b_R}$ larger than $0.5$ is not acceptable, as it would lead to a bottom mass $m_b \lesssim 10^{-2}$ GeV, so the RS--GIM
mechanism cannot be fully effective.
\\ In the present context of our mechanism mixing $b^{(0)}_{1L}$ and $b^{(0)}_{2L}$, the situation is improved. Indeed,
we can choose the parameter controlling the bottom mass: $c_2$ [see
Eq.(\ref{5Daction})] to be smaller than $0.5$ (as done in
Fig.(\ref{fig:MODI})) in order to generate a correct $m_b$ value. The important point here is that the bottom quark is mainly composed by the field $b^{(0)}_{1L}$ (see the 
$\cos \theta$ value in Fig.(\ref{fig:MODI})) to which is affected a {\it different} $c$--parameter: $c_1$. Hence, we can fix $c_1$ independently
from $c_2$ and a priori take $c_1$ larger than $0.5$ to optimize the RS--GIM
mechanism. Concretely, $c_1$ can reach values up to $\sim 0.5$ [see  
Fig.(\ref{fig:MODI}) where remarkably $c_1$ is larger than the
relevant $c_2$ values, illustrating the improvement with respect to the usual
case of a common $c_{Q_L}$] 
due to the additional requirement of solving the $A^b_{FB}$ anomaly (a large $c_1$ can give rise to too small 
$\vert \delta g_{Z^0}^{b_{L/R}}/g_{Z^0}^{b_{L/R}} \vert_{{\rm TOTAL}} \vert$). 
\\ The FCNC effect amplitudes depend crucially on global three flavor mass matrices and thus on parameter values for each SM fermion generation. 
Since our goal here is not to elaborate a complete three flavor model (predicting all Yukawa couplings and $c$--parameters), as already mentioned, 
we will not compute the precise FCNC effects in the down--quark sector.
Notice that the calculation of the contribution to the B mass splitting from KK gluon exchange must also include $b-b'$ mixing effects,
which has not been done in details so far to our knowledge.  
\\ Anyway, several recent works \cite{FlavSym} show that FCNC processes can be suppressed by gauging some (SM) flavor symmetries in the bulk,
precisely like the SM custodial symmetry is gauged in the bulk to reduce corrections on EW observables \cite{FlavRandall}. An example of
flavor symmetry in this context is the non--abelian discrete symmetry $A_4$ \cite{Grojean}. 
\\ \\
\noindent {\bf Fit of $A^b_{FB}$ and $R_b$:}
Now that the way of calculating the shifts $\delta g_{Z^0}^{b_{L/R}}/g_{Z^0}^{b_{L/R}} \vert_{{\rm TOTAL}}$ with respect to SM has been described, 
one is in a position to compute the predictions for values of $A^b_{FB}$ \footnote{Note that $b-b'$ type mixings do not give rise to corrections
of the photon coupling, involved in the $A^b_{FB}$ computation, since $b$ and $b'$ have the same electric charge. Besides, there is no mass mixing 
between the photon and its KK excitations.} and $R_b$ in the RS context (like in \cite{RSAFB}). 
In Fig.(\ref{fig:MODI}) we present contour plots of the p--value in the plan $\{c_{t_R},c_2\}$ resulting from the global fit of $R_b$ plus the eight
measurements of $A^b_{FB}$: three in the $Z^0$ pole energy region \cite{PDG,LEPex}, four at center of mass energies below \cite{AFB-below,Klaus,AFB-mix} 
and the last one far above ($190.7$ GeV) \cite{AFB-above}. 
In the same figure, we show contour plots of bottom and top quark masses. The bottom mass $m_b$ is obtained after bi--diagonalizing numerically the mass
matrix (\ref{MbTOTAL}) [{\it c.f.} Eq.(\ref{MbEIGEN})]. The top mass $m_t$ is obtained similarly once an analog top quark mass matrix is derived.

We now discuss the choice of parameter values taken in Fig.(\ref{fig:MODI}) starting with the Yukawa coupling constants which are dimensionful, indicative of the
non--renormalizable nature of the 5--dimensional theory. We fix the cut--off of the effective field theory at $\Lambda_{IR} \simeq 2 M_{KK}$ since the two heaviest eigenstates considered,
mainly composed by the $b^{c\prime(2)}$ and $b^{c\prime\prime(2)}$ fields, can reach mass values at most equal to $2 M_{KK}$ typically [as $c_{t_R}$, which determines $m^{\prime(2)}_{c_{t_R}}$,
remains smaller than $0.2$ in Fig.(\ref{fig:MODI})]. This choice thus allows to trust in our treatment of the Higgs coupling to bottom quarks on which relies the mass matrix (\ref{MbTOTAL}). 
Indeed, up to the energy scale $\Lambda_{IR}$, the proper effective dimensionless Yukawa coupling constant localized on the TeV--brane $\Lambda \lambda_b^{IR}$ is required to be in the perturbative 
regime: $(\Lambda_{IR} \lambda_b^{IR})^2 / 16 \pi^2 \lesssim 1$.
This condition of a weakly coupled theory expresses the fact that typical loop corrections are smaller than the lowest order terms. 
The brane localized Yukawa coupling is related to the bulk Yukawa coupling [entering Eq.(\ref{5Daction})] (used previously e.g. in \cite{RSloc,ADMS}) through $\lambda_b^{IR} k e^{- \pi k R_c} = \lambda_b^{5D} k$ 
(or $\lambda_b^{IR} / R' = \lambda_b^{5D} / R$ using current notations), as confirmed by comparing the zero--mode mass in Eq.(\ref{MbTOTAL}) and in recent references e.g. \cite{CFW,FlavSym,Grojean}.
Hence, the perturbativity condition leads to $\vert \lambda_b^{5D} \vert k \lesssim 3$ (after replacing by the $M_{KK}$ expression). 
We also impose $\vert \lambda_b^{5D} \vert k \gtrsim k / M_5$, $M_5$ being the fundamental 5--dimensional Planck scale entirely fixed by the $kR_c$ and $k$ (or equivalently $M_{KK}$) values \cite{RS}. 
For the $M_{KK}$ values considered ($3$,$4$,$5$ TeV), one gets $\vert \lambda_b^{5D} \vert k \gtrsim 0.1$. 
These upper and lower limits on $\vert \lambda_b^{5D} \vert k$ guarantee that no totally new energy scale value associated to $\vert \lambda_b^{5D} \vert^{Ñ1}$ is introduced in the theory. 
Similar constraints are derived for $\vert \lambda_t^{5D} \vert k$. The values for $\vert \lambda_{b,t}^{5D} \vert k$ considered in Fig.(\ref{fig:MODI}) satisfy all these constraints.
Furthermore, these $\vert \lambda_{b,t}^{5D} \vert k$ values are chosen such that the parameters $c_{b_R}$ and $c_1$ are maximized, which allows to minimize FCNC effects (see above
discussion). 
\\ Concerning the mass $M_{KK}$ and coupling $g_{Z'}$ as well as the sign entering the $\sin^2 \theta'$ expression (\ref{thetaP}), in Fig.(\ref{fig:MODI}) we take them exactly as
in the estimation of the set of three theoretical predictions for $S_{\rm RS}$,$T_{\rm RS}$ in Fig.(\ref{fig:RS}). In particular, the sign in $\sin^2 \theta'$ is negative and the coupling
lies inside the allowed range $0.71 \lesssim g_{Z'} \lesssim 2.23$. For
consistency, it is important to consider identical values for these three common parameters entering both the $S_{\rm RS}$,$T_{\rm RS}$ analysis of Section \ref{conf} and the present
EW analysis of the heavy generation sector. 
\\ Finally, we have maximized the $c_{b_R}$ and $c_1$ values in order to to minimize FCNC effects. It means that we have chosen the largest $c_{b_R}$ and $c_1$ values
such that it still exist a common region in the plan $\{c_{t_R},c_2\}$ where {\it (i)} reasonable $m_{b,t}$ values are reproduced {\it (ii)} the EW fit is improved compared to the SM case.
Without maximizing $c_{b_R}$ and $c_1$, the best--fit curves in Fig.(\ref{fig:MODI}) at $M_{KK}=5$ TeV could have been centered with respect to the domain of acceptable $m_{b,t}$.

\begin{figure}[!ht]
\begin{center}
\includegraphics[width=7.5cm]{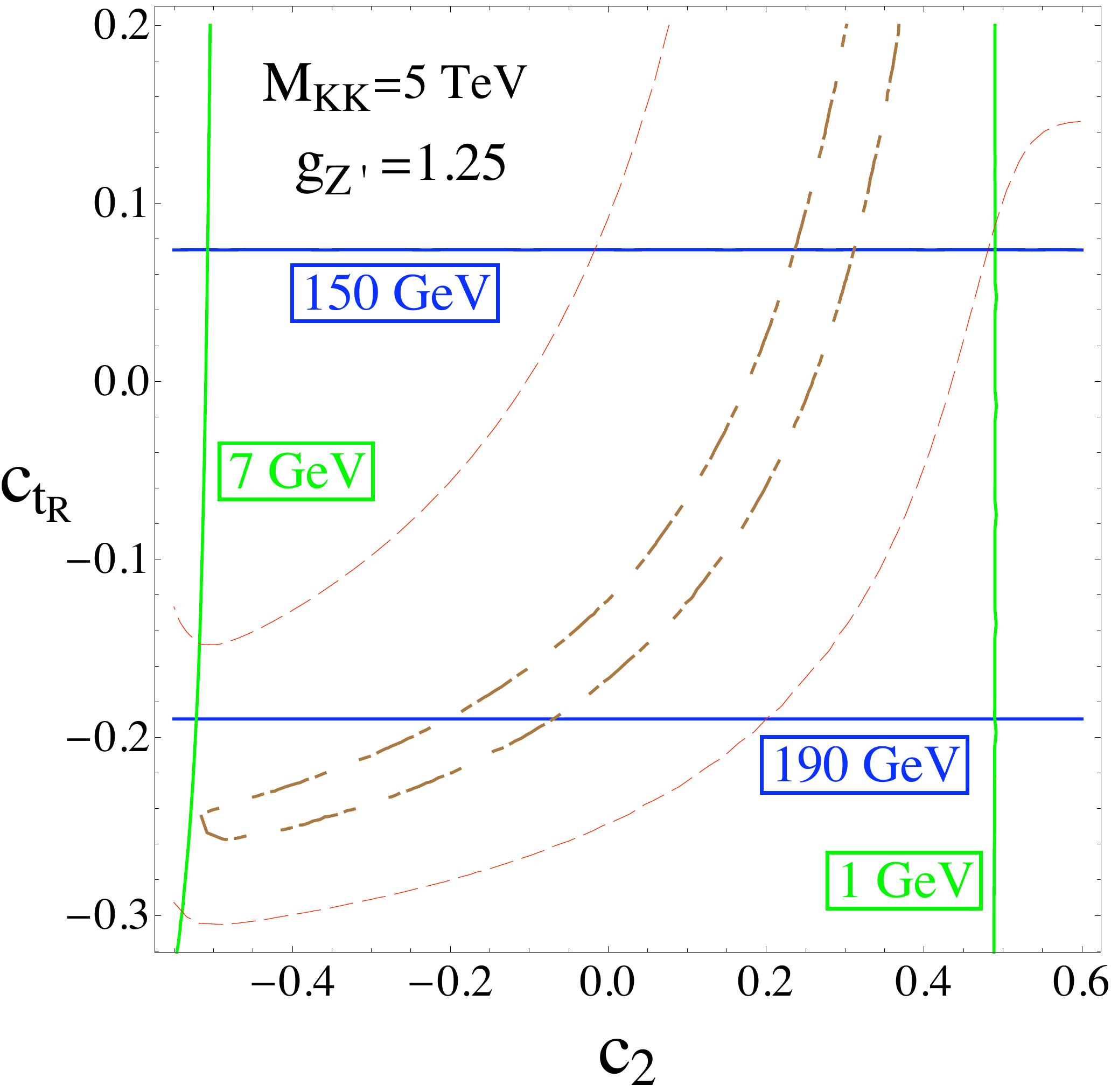} 
\end{center}
\vspace*{-5mm}
\caption{Contour levels in the plan $\{c_{t_R},c_2\}$ at $28.3 \%  $ [$\chi^2/d.o.f. \equiv 1.21$ with a degree of freedom at $9$] (brown dot--dashed--lines)
and $10 \%  $ [$\chi^2/d.o.f. \equiv 1.63$] (red dashed--lines) for the fit of the experimental data on $R_b$ and the eight measurements of $A^b_{FB}$,
with $m_h = 500$ GeV. For comparison, the SM fit reaches $27.6 \%  $ [$\chi^2_{SM}/d.o.f. \equiv 1.22$]. Quark masses are also indicated as previously.  
The plot is obtained for Model I with $\cos^2\theta=0.11$, $\lambda_{t}^{5D} k=3$, $\lambda_{b}^{5D} k=0.15$, $c_{b_R}=0.35$, $c_1=0.45$, 
$g_{Z'}=1.25$ and $M_{KK}=5$ TeV.}
\label{fig:MODI500}
\end{figure}

As already described, the produced SM quark masses increase as their $c$--parameters decrease (which localizes them towards the TeV--brane where lives the Higgs boson). 
This manifests itself in the expression for the zero--mode mass given in first line and first column of matrix (\ref{MbTOTAL}) in case of the bottom quark.
For example, we see clearly on Fig.(\ref{fig:MODI}) that $m_b$ increases as $c_2$ decreases, $c_2$ being the parameter controlling the bottom mass
(see Eq.(\ref{5Daction})). The expression for the zero--mode mass in Eq.(\ref{MbTOTAL}) shows also that $m_b$ is suppressed by small $\cos \theta$ 
values, as it occurs in Fig.(\ref{fig:MODI}). Indeed, these small values mean that the bottom quark is mainly composed by the $b^{(0)}_{1L}$ which has no pure zero--mode
mass term.
\\ The choice of letting such an uncertainty on theoretical predictions for $m_{b,t}$ in Fig.(\ref{fig:MODI}) (respective ranges, in GeV, of $[1,7]$ and $[150,190]$) 
and of not trying to reproduce exactly the experimental $m_{b,t}$ values
is justified by the three flavor mixing effect (see above) together with the following considerations on energy. 
The cut--off energy scale at which the extra--dimension is integrated out is typically equal to the lightest KK mass which can be of few hundreds of GeV only (as 
discussed later for custodians). In calculation of predicted values for the EW observables: 
$R_b$, $A^b_{FB}$, $m_{W^\pm}$, $\Gamma_{\ell \ell}$ and $\sin^2\theta_{{\rm eff}}^{{\rm lept}}$, we have considered $g_{Z'}$ as well as the 4--dimensional effective 
dimensionless Yukawa coupling $\lambda_{b}^{5D} k$ (up to geometrical factors) at the typical $Z^0$ pole energy scale. Hence, the experimental values for $m_{b,t}$ to be reproduced
must also be considered at an energy $\mu \simeq m_{Z^0}$. The effect of renormalization group from the pole masses to the $Z^0$ boson mass scale can be important. 
For instance, within the pure SM, this effect is of $29.5\%$ on the bottom mass which is $m_b=4.248 \pm 0.046$ GeV at its own pole mass, 
and the same effect reaches a correction of $6.5\%$ for $m_t$ (still in the SM) \cite{FusKo}. An updated rigorous computation of these corrections within the RS context,
which relies on loop diagrams including effects from KK states, is beyond the scope of the present work. We thus allow for some relative uncertainty on the 
$m_{b,t}$ values at $\mu \simeq m_{Z^0}$.
\\ \\
\noindent {\bf Discussion:}
Let us draw our conclusion. One deduces from Fig.(\ref{fig:MODI}) that, in the Model I and for $M_{KK}=3,5$ TeV, there are allowed regions of the complete parameter space where:
the global fit of $R_b$ and $A^b_{FB}$ [at the various center of mass energies] is significantly improved with respect to the analog fit of $R_b$ and $A^b_{FB}$ performed within the 
pure SM situation (which is only at $0.8 \%  $), and simultaneously, the experimental $m_{b,t}$ values are potentially reproducable. We also mention in particular that 
in obtained regions where the EW fit is typically better than $10 \%  $, the difference between the theoretical expectation for $A^b_{FB}(m_{Z^0})$ and the corresponding
measured value is below one standard deviation, thus solving the $A^b_{FB}$ anomaly at the $Z^0$ pole. 
\\ Furthermore, we observe that in relevant regions of parameter space obtained (where EW fit is improved and $m_{b,t}$ are reproduced), custodian fermions can reach low masses. 
For example, the first KK excitation of exotic quark $q^{c\prime}_{{\rm (5/3)}}$ (in the $\{t^c_R\}$ multiplet) gets a mass $m^{\prime(1)}_{c_{t_R}}=1169$ GeV
[there is no mass contribution originating from Yukawa couplings] at the point of low $c_{t_R}$: $c_{t_R}=-0.30$, $c_2=0.17$ in the plot of Fig.(\ref{fig:MODI})
with $M_{KK}=3$ TeV \footnote{$m^{\prime(1)}_{c_{t_R}}$ depends solely on $kR_c$, $c_{t_R}$ and $M_{KK}$.}.
\\ \\
\noindent {\bf Heavy Higgs boson regime:}
We finish this part by discussing the Higgs mass dependence, a discussion which holds also for the models considered in next two subsections as similar conclusions can be established. 
In Fig.(\ref{fig:MODI}) where are presented results of the fit between the experimental data on $R_b$,$A^b_{FB}$ and their theoretical prediction (including the RS--induced corrections),
the theoretical SM expectations for these EW observables \cite{LEPex} are taken at the reference value for the Higgs mass: $m_h=100$ GeV \cite{RSAFB} (similarly, e.g. $m_{W^\pm} \vert_{SM}$
depends on the Higgs mass as discussed in Section \ref{conf}).
The variation of these theoretical SM values with $m_h$ is given in a good approximation \cite{Neubert} by,
\begin{eqnarray}   
\Delta R_b \ = \ 3.3 \ 10^{-5} \ \mbox{ln} \bigg ( \frac{m_h}{m_h^{ref}} \bigg ) \ ;
\ \ \
\Delta A^b_{FB}(m_{Z^0}) \ = \ -2.7 \ 10^{-3} \ \mbox{ln} \bigg ( \frac{m_h}{m_h^{ref}} \bigg ) ,
\label{VARHiggs}   
\end{eqnarray}
where we have chosen the Higgs mass reference value to be $m_h^{ref}=100$ GeV.
It is clear from Eq.(\ref{VARHiggs}) that increasing $m_h$ from $100$ GeV up to e.g. $115$ GeV would not lead to large
corrections of the theoretical SM predictions for $R_b$ and $A^b_{FB}(m_{Z^0})$, 
relatively to their experimental error (see beginning of Section \ref{Third}).
In contrast, an heavy Higgs such that $m_h \gtrsim 500$ GeV leads to an important decrease of the SM expectation for $A^b_{FB}$ at the $Z^0$ pole,
as well as at the two other LEP energies: $A^b_{FB}(89.55\mbox{GeV})$ and $A^b_{FB}(92.94\mbox{GeV})$ [assuming the same dependence on $m_h$],
allowing to address the anomaly around the pole (compare in particular the new SM prediction of $A^b_{FB}(m_{Z^0})$ with its experimental data 
given at the beginning of Section \ref{Third}) since $R_b$ is not significantly affected by this $m_h$ increase and thus remains in a good
agreement with its experimental estimation. 
Then, the corrections $\delta g_{Z^0}^{b_{L/R}}/g_{Z^0}^{b_{L/R}} \vert_{{\rm TOTAL}}$ due to KK state mixings can still help to improve the 
SM fit of $R_b$ and $A^b_{FB}$ (at the various energies) data but no more in a significant way, as shown in Fig.(\ref{fig:MODI500}). Indeed
with $m_h = 500$ GeV, the SM fit is at $27.6 \%  $ while the RS fit reaches at most $29.0 \%  $ [e.g. at the point $c_2=0.18$, $c_{t_R}=0$]
for the example of parameter set described in Fig.(\ref{fig:MODI500}).
\\ It was shown in \cite{Neubert} that if the Higgs boson is heavy, e.g. $m_h = 400$ GeV, the $A^b_{FB}(m_{Z^0})$ anomaly  
could also be addressed without custodial symmetry in the bulk. In such a scenario, the only corrections to the $Z^0 \bar b b$ vertex
would come from the mixings of the $Z^0$ boson and $b$ quark with their respective KK excitations.   
The smallest KK mass would have to be larger than $\sim 4$ TeV in order to soften the corrections 
to the SM predictions for $A^b_{FB}$ and $R_b$ which would already be in good agreement with their experimental measurement.

\subsection{Model II}

The second model is characterized by the representations:
\begin{eqnarray}
\{Q_{1L}\} \ \equiv \ 
({\bf 2},{\bf 3})_{1/6} \ = \
\left(\begin{array}{ccc} q^{\prime}_{{\rm (5/3)} L} & t_{1L} & b'_L \\ t'_L & b_{1L} & q^{\prime}_{{\rm (-4/3)} L} \end{array}\right) 
\ \ \ 
\{t^c_R\} \ \equiv \ 
({\bf 3},{\bf 2})_{1/6} \ = \
\left(\begin{array}{cc} q^{c\prime}_{{\rm (5/3)} R} & t^{c\prime}_R \\ t^c_R & b^{c\prime\prime}_R \\ b^{c\prime}_R & q^{c\prime}_{{\rm (-4/3)} R} \end{array}\right)
\label{repII:Q1LtcR}
\end{eqnarray}
\begin{eqnarray}
\{Q_{2L}\} \ \equiv \ 
({\bf 2},{\bf 3})_{-5/6} \ = \
\left(\begin{array}{ccc} t_{2L} & b''_L & q'''_{{\rm (-4/3)} L} \\ b_{2L} & q''_{{\rm (-4/3)} L} & q'_{{\rm (-7/3)} L} \end{array}\right) 
\ \ \ 
\{b^c_R\} \ \equiv \ 
({\bf 1},{\bf 2})_{-5/6} \ = \
\left(\begin{array}{cc} b^c_R & q^{c\prime\prime}_{{\rm (-4/3)} R} \end{array}\right)
\label{repII:Q2LbcR}
\end{eqnarray}

Here we do not present in details the mass matrices and the derivation of $Z^0$ couplings to bottom quarks in the mass basis which allow to determine 
$\delta g_{Z^0}^{b_{L/R}}/g_{Z^0}^{b_{L/R}} \vert_{{\rm TOTAL}}$, as the method is similar.
\\ In the present model, $\delta g_{Z^0}^{b_R^c} / g_{Z^0}^{b_R^c} \vert_{{\rm boson}}>0$ and $\delta g_{Z^0}^{b_R^c} / g_{Z^0}^{b_R^c} \vert_{{\rm fermion}}>0$ 
which lead to $\delta g_{Z^0}^{b_{R}^c}/g_{Z^0}^{b_{R}^c} \vert_{{\rm TOTAL}} > 0$, whereas
$\delta g_{Z^0}^{b_L} / g_{Z^0}^{b_L} \vert_{{\rm boson}} <0$ and $\delta g_{Z^0}^{b_L} / g_{Z^0}^{b_L} \vert_{{\rm fermion}} >0$ 
giving $\delta g_{Z^0}^{b_{L}}/g_{Z^0}^{b_{L}} \vert_{{\rm TOTAL}} < 0$.

\begin{figure}[!ht]
\begin{center}
\includegraphics[width=7.5cm]{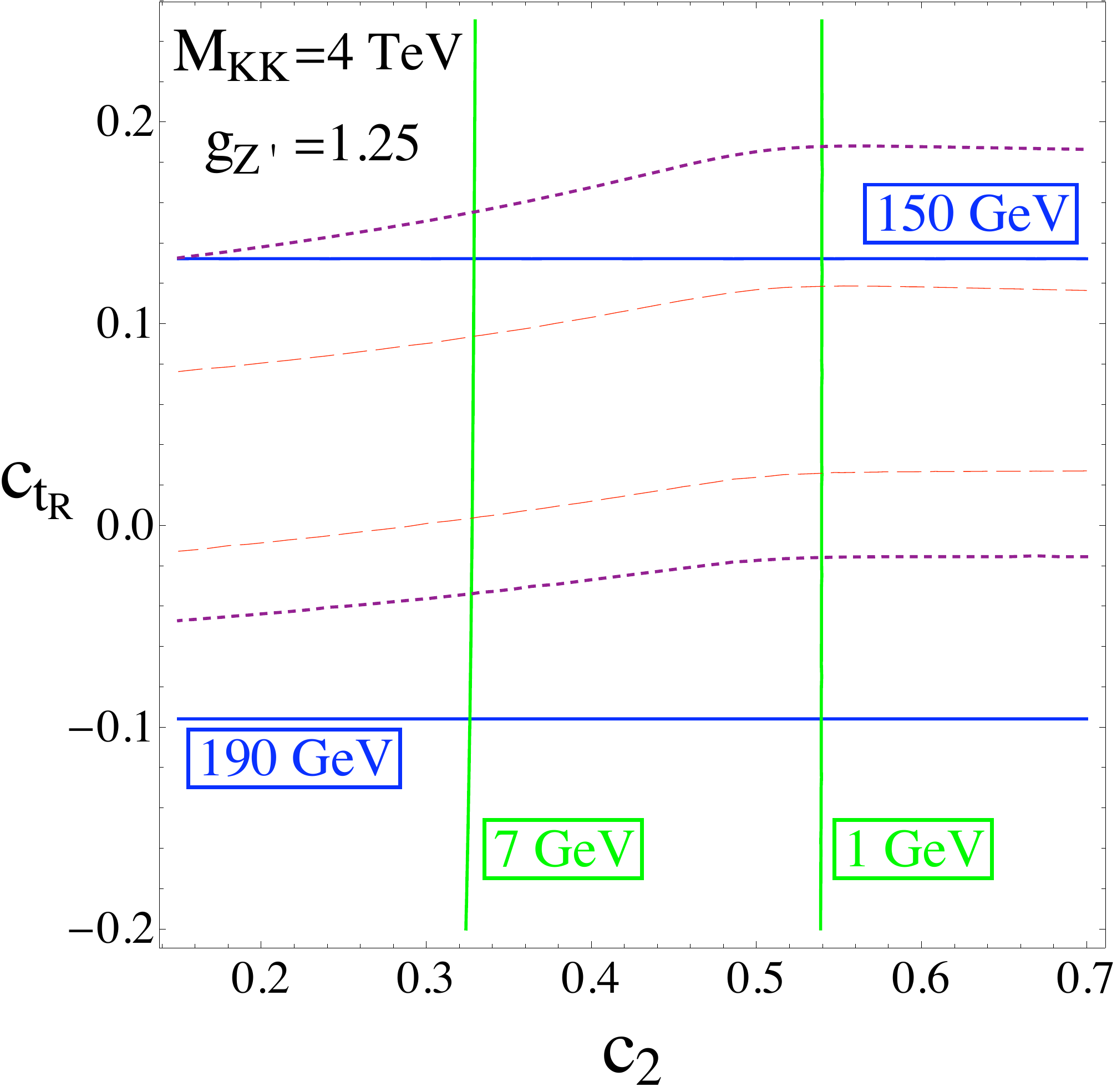} 
\end{center}
\vspace*{-5mm}
\caption{Contour levels in the plan $\{c_{t_R},c_2\}$ at $10 \%  $ [$\chi^2/d.o.f. \equiv 1.63$ with a degree of freedom at $9$] (red dashed--lines)
and $8 \%  $ [$\chi^2/d.o.f. \equiv 1.71$] (purple dotted--lines) for the fit of the experimental data on $R_b$ and the eight measurements of $A^b_{FB}$ at different
center of mass energies. For comparison, in the SM, this fit reaches only $0.8 \%  $ [$\chi^2_{SM}/d.o.f. \equiv 2.47$].
We also present the curves corresponding to $m_b=1;7$ GeV (green plain--lines) and $m_t=150;190$ GeV (blue plain--lines).  
This plot is for Model II with $M_{KK}=4$ TeV, $\cos^2\theta=0.01$, $\lambda_{t}^{5D} k=1$, $\lambda_{b}^{5D} k=1$, $c_{b_R}=0.27$, $c_1=0$.}
\label{fig:MODII}
\end{figure}

In Fig.(\ref{fig:MODII}) are shown for Model II the contour plots of the p--value, for the fit of $R_b$ and the eight
measurements of $A^b_{FB}$, together with contour plots of bottom and top quark masses. 
The results and contour plots for this model are generically very close to those obtained for the previous one: Model I.
Indeed, the main difference between these two models is the appearance of a state $b'_L$ in the $\{Q_{1L}\}$ multiplet ({\it c.f.} Eq.(\ref{repII:Q1LtcR}))
but without direct mixing mass term of the type $\bar b^{c(0)}_R b^{\prime(1)}_L$. By consequence, no significant mixing of the SM $b$ quark with this new $b'$ arises.
Besides, here the quantum numbers of $t_{1L}$ and $b_{1L}$ with respect to the custodial symmetry (especially $I_{3R}^{b_{1L}}$) are not modified relatively to Model I
so that $Q_{Z'}^{b_{1L}}$ [entering Eq.(\ref{dgsgBOSON})] is unchanged.
\\ Due to this similarity of Models I and II, we have chosen to illustrate here other domains of parameter space (than in Fig.(\ref{fig:MODI})) 
where the fit of $R_b+A^b_{FB}$ is improved with respect to SM and $m_{b,t}$ are reproducable. For example, in Fig.(\ref{fig:MODII}) we have chosen 
to consider the intermediate case of $M_{KK}=4$ TeV and $g_{Z'}=1.25$. The effective Yukawa coupling constants have also been fixed at unity,
$\lambda_{t}^{5D} k=1$ and $\lambda_{b}^{5D} k=1$, to demonstrate that interesting regions are accessible without adjusting these couplings.
Notice that, in Fig.(\ref{fig:MODII}), all coupling constants, namely $\lambda_{t}^{5D}$, $\lambda_{b}^{5D}$ and $g_{Z'}$, have values which 
do not saturate their perturbative limit. 
In particular, as $\lambda_{b}^{5D} k=1$, the cut--off can even be enhanced to $\Lambda_{IR} \simeq 5 M_{KK}$ which is still compatible with
$(\Lambda_{IR} \lambda_b^{IR})^2 / 16 \pi^2 \lesssim 1$.

\subsection{Model III}

The last scenario is defined by,
\begin{eqnarray}
\{Q_{1L}\} \ \equiv \ 
({\bf 2},{\bf 2})_{2/3} \ = \
\left(\begin{array}{cc} q^{\prime}_{{\rm (5/3)} L} & t_{1L} \\ t^{\prime}_L & b_{1L} \end{array}\right) 
\ \ \ 
\{t^c_R\} \ \equiv \ 
({\bf 1},{\bf 3})_{2/3} \ = \
\left(\begin{array}{ccc} q^{c\prime}_{{\rm (5/3)} R} & t^c_R & b^{c\prime}_R \end{array}\right)
\label{repIII:Q1LtcR}
\end{eqnarray}
\begin{eqnarray}
\{Q_{2L}\} \ \equiv \ 
({\bf 2},{\bf 3})_{-5/6} \ = \
\left(\begin{array}{ccc} t_{2L} & b'_L & q''_{{\rm (-4/3)} L} \\ b_{2L} & q'_{{\rm (-4/3)} L} & q'_{{\rm (-7/3)} L} \end{array}\right) 
\ \ \ 
\{b^c_R\} \ \equiv \ 
({\bf 1},{\bf 2})_{-5/6} \ = \
\left(\begin{array}{cc} b^c_R & q^{c\prime}_{{\rm (-4/3)} R} \end{array}\right)
\label{repIII:Q2LbcR}
\end{eqnarray}

For these group representations,
one has $\delta g_{Z^0}^{b_R^c} / g_{Z^0}^{b_R^c} \vert_{{\rm boson}}>0$, $\delta g_{Z^0}^{b_R^c} / g_{Z^0}^{b_R^c} \vert_{{\rm fermion}}>0$ 
and $\delta g_{Z^0}^{b_{R}^c}/g_{Z^0}^{b_{R}^c} \vert_{{\rm TOTAL}} > 0$, while
$\delta g_{Z^0}^{b_L} / g_{Z^0}^{b_L} \vert_{{\rm boson}} >0$, $\delta g_{Z^0}^{b_L} / g_{Z^0}^{b_L} \vert_{{\rm fermion}} <0$ 
and $\delta g_{Z^0}^{b_{L}}/g_{Z^0}^{b_{L}} \vert_{{\rm TOTAL}} < 0$.

\begin{figure}[!ht]
\begin{center}
\includegraphics[width=7.cm]{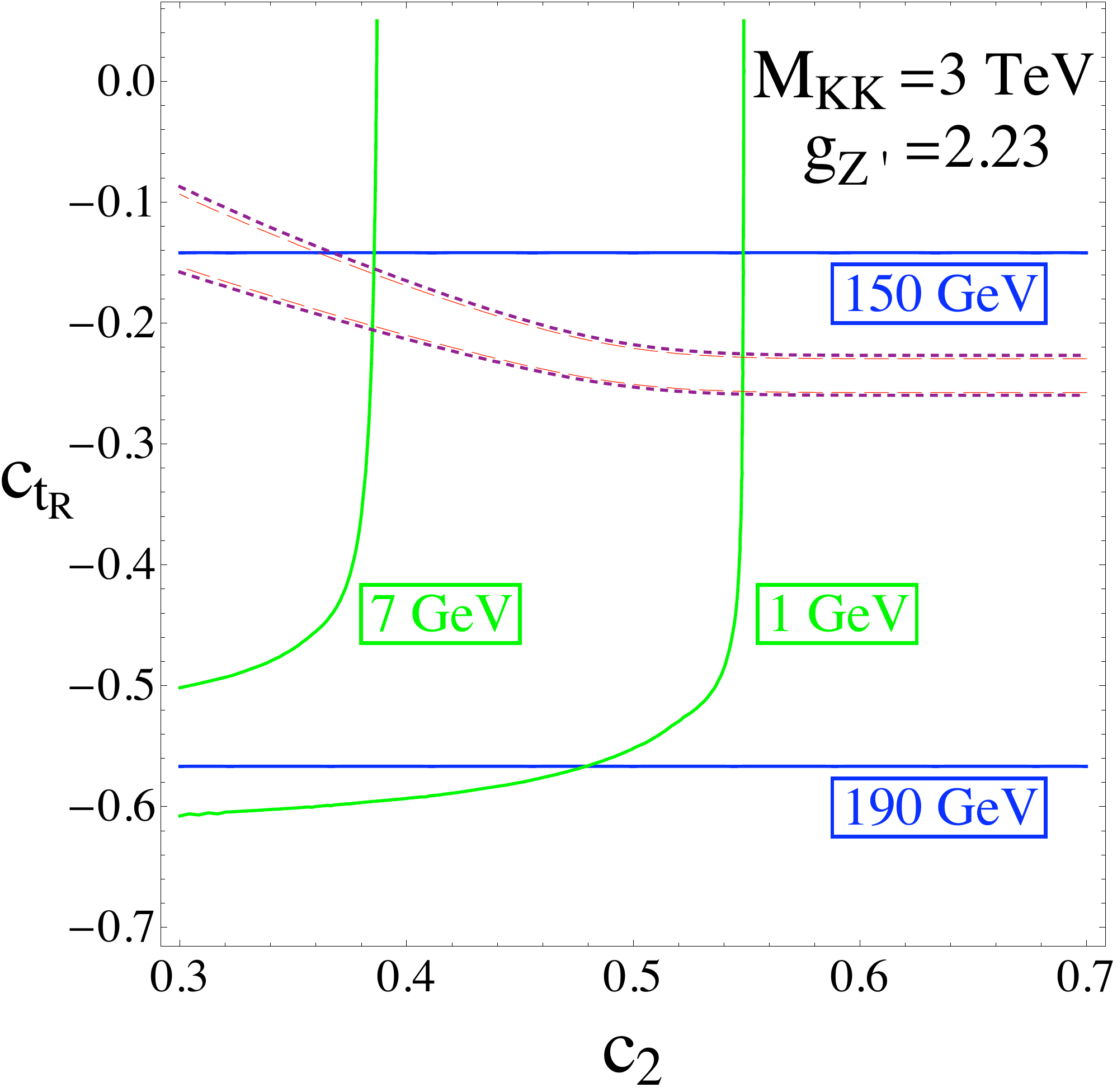} 
\includegraphics[width=7.cm]{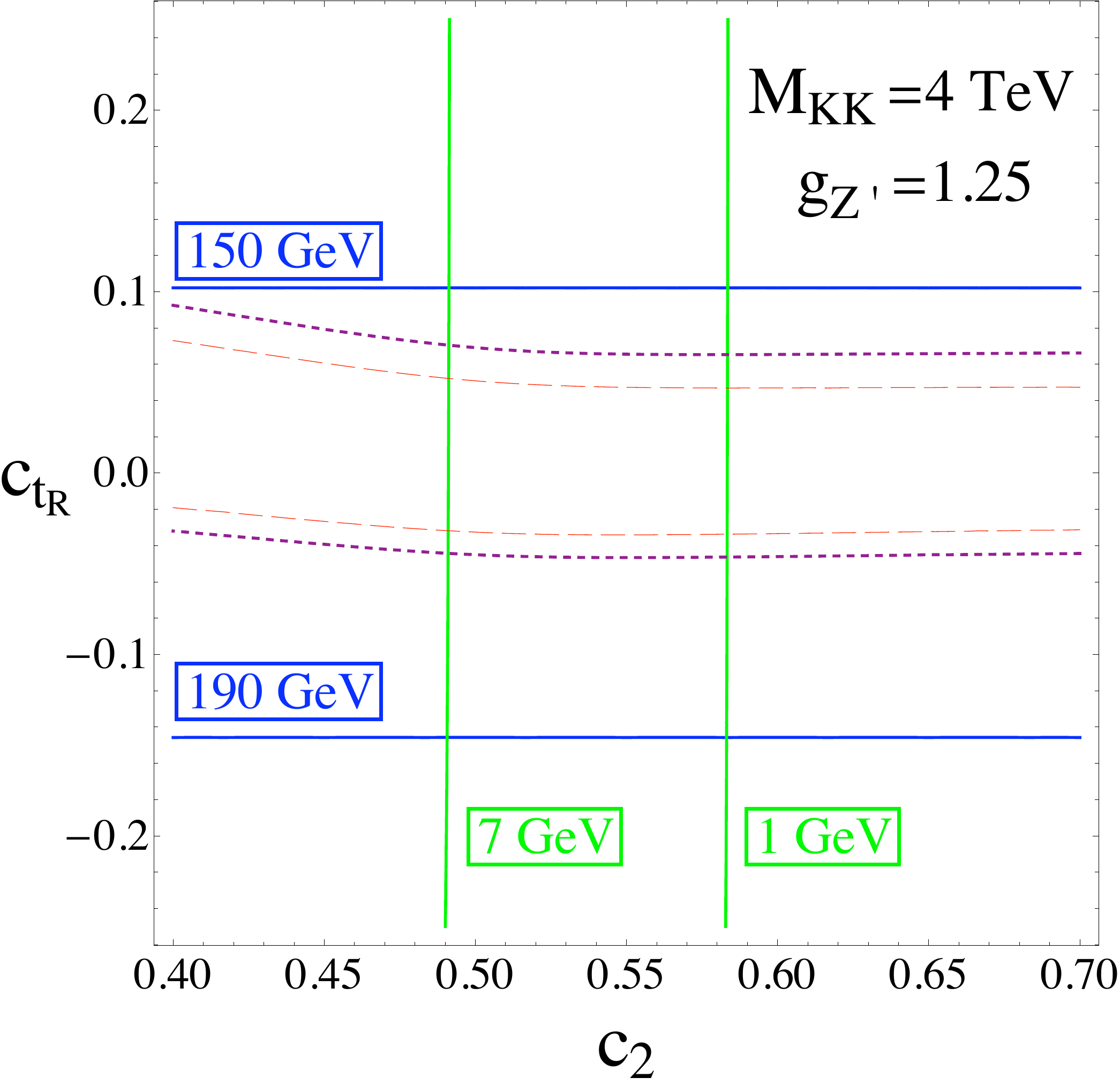} 
\includegraphics[width=7.cm]{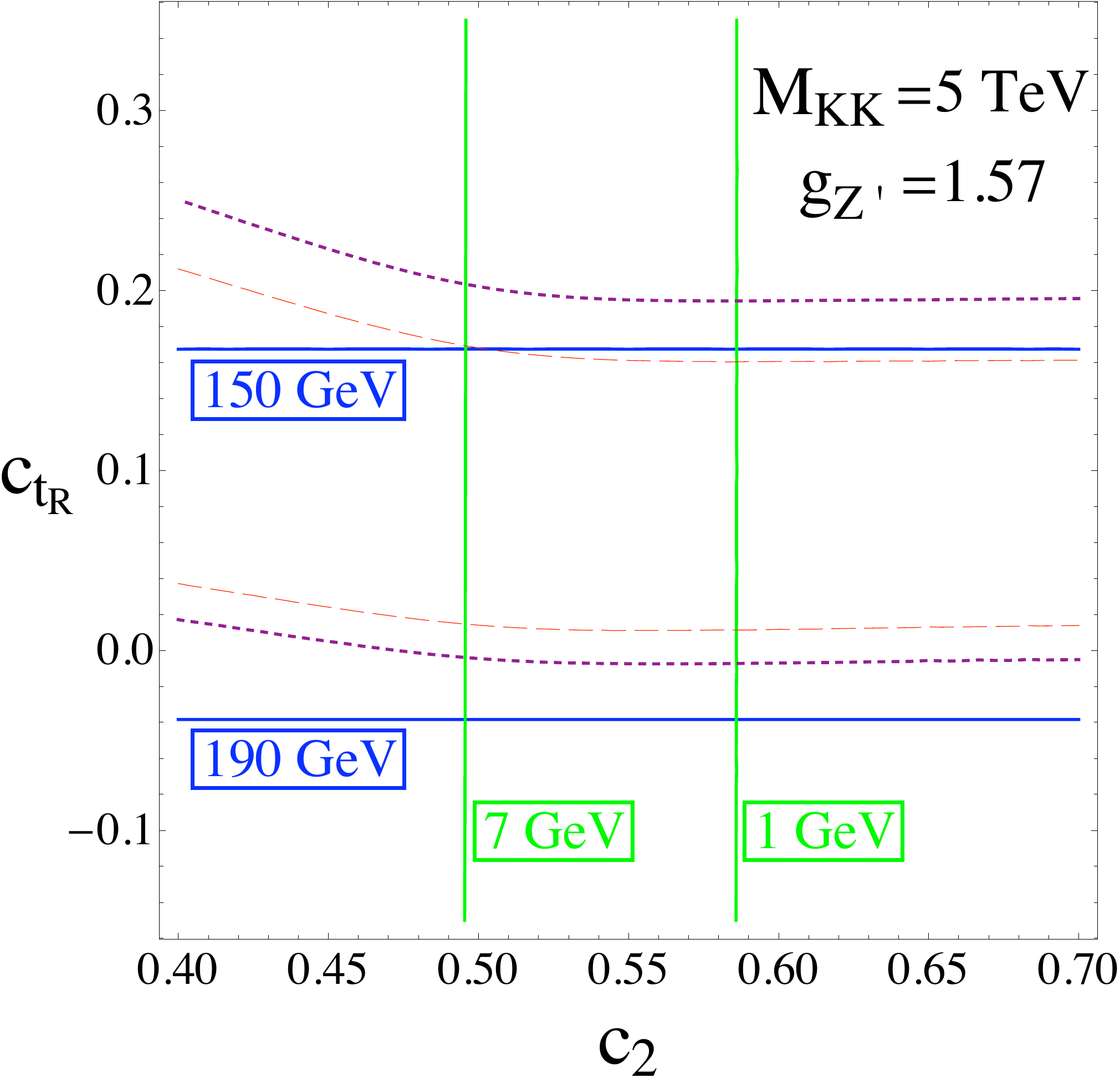} 
\end{center}
\vspace*{-5mm}
\caption{Contour levels in the plan $\{c_{t_R},c_2\}$ at $10 \%  $ [$\chi^2/d.o.f. \equiv 1.63$ with a degree of freedom at $9$] (red dashed--lines)
and $8 \%  $ [$\chi^2/d.o.f. \equiv 1.71$] (purple dotted--lines) for the fit of the experimental data on $R_b$ and the eight measurements of $A^b_{FB}$ at different
center of mass energies. For comparison, in the SM, this fit reaches only $0.8 \%  $ [$\chi^2_{SM}/d.o.f. \equiv 2.47$].
Also presented are the curves for $m_b=1;7$ GeV (green plain--lines) and $m_t=150;190$ GeV (blue plain--lines).  
These three plots, associated to different values of $g_{Z'}$ and $M_{KK}$ (as indicated on the figure), are obtained for Model III with 
$\cos^2\theta=0.05$, $\lambda_{t}^{5D} k=2.1$, $\lambda_{b}^{5D} k=1.0$, $c_{b_R}=0.45$, $c_1=0.45$ [plot for $M_{KK}=3$ TeV]; 
$\cos^2\theta=0.3$, $\lambda_{t}^{5D} k=1.65$, $\lambda_{b}^{5D} k=1.60$, $c_{b_R}=0.25$, $c_1=0.33$ [plot for $M_{KK}=4$ TeV]; 
$\cos^2\theta=0.3$, $\lambda_{t}^{5D} k=2.8$, $\lambda_{b}^{5D} k=1.7$, $c_{b_R}=0.25$, $c_1=0.43$ [plot for $M_{KK}=5$ TeV].}
\label{fig:MODIII}
\end{figure}

In Fig.(\ref{fig:MODIII}), we present the contour plots for the p--value as well as for bottom and top quark masses, in Model III. 
At $M_{KK}=3$ TeV, the $c_{t_R}$ typical parameter values are minimized for reducing the KK mass of the $b^{c\prime(1)}$ mode (in the $\{t^c_R\}$ multiplet). 
For example, at the possible point $c_{t_R}=-0.25$, $c_2=0.50$, this mass is $m^{\prime(1)}_{c_{t_R}}=1320$ GeV giving rise to a second mass eigenvalue $m_{b_2}=1241$ GeV 
(see Eq.(\ref{MbEIGEN})). This $c_{t_R}$ minimization is not going with an optimum centering of p--value lines as shows Fig.(\ref{fig:MODIII}).
\\ Concerning the two plots of Fig.(\ref{fig:MODIII}) at $M_{KK}=4,5$ TeV, $c_{b_R}$ and $c_1$ have been maximized to reduce FCNC effects.

For instance, at the point $c_{t_R}=0$, $c_2=0.55$ of Fig.(\ref{fig:MODIII}) for $M_{KK}=4$ TeV,
one has e.g. $\delta g_{Z^0}^{b_L} / g_{Z^0}^{b_L} \vert_{{\rm boson}} \simeq + 1.73\%$ and $\delta g_{Z^0}^{b_L} / g_{Z^0}^{b_L} \vert_{{\rm fermion}} \simeq - 1.72\%$ 
which illustrates the fact that the fermion mixing effect is not neglectable at all compared to the boson one; it was important to consider this
fermion effect throughout this paper.

As a conclusion, it is also possible in this Model III to find some regions of the parameter space, with $M_{KK}=3,4,5$ TeV, where both conditions of EW fit improvement and mass
reproduction can be satisfied. For instance, from the plot of Fig.(\ref{fig:MODIII}) for $M_{KK}=3$ TeV, we find that the EW fit reaches $17.3 \%  $ [$\chi^2/d.o.f. \equiv 1.42$] 
in the RS scenario at the point $c_{t_R}=-0.24$, $c_2=0.53$.

\subsection{Other models}
\label{subO3}

\noindent {\bf Minimal model:}
The first model we comment on in this last subsection is called minimal since it
possesses the minimum field content, and, it is the one which was originally 
proposed when the custodial symmetry was introduced in the RS scenario
\cite{ADMS}. The minimal model is characterized by these quark representations:
\begin{eqnarray}
\{Q_{L}\} \ \equiv \ 
({\bf 2},{\bf 1})_{1/6} \ = \
\left(\begin{array}{c} t_{L} \\ b_{L} \end{array}\right) 
\ \ \ 
\{t^c_R\} \ \equiv \ 
({\bf 1},{\bf 2})_{1/6} \ = \
\left(\begin{array}{cc} t^{c}_R & b^{c\prime}_{R} \end{array}\right)
\label{Min:Q1LtcR}
\end{eqnarray}
\begin{eqnarray}
\{b^c_R\} \ \equiv \ 
({\bf 1},{\bf 2})_{1/6} \ = \
\left(\begin{array}{cc} t^{c\prime}_R & b^{c}_{R} \end{array}\right)
\label{Min:Q2LbcR}
\end{eqnarray}
The authors of \cite{LRarg} have shown that, for $\tilde g=g$ [a choice
which fixes $g_{Z'}$ as shows Eq.(\ref{thetaP})], $M_{KK}=3.75$ TeV 
and values of $c_{Q_L}$,$c_{t_R}$ reproducing the top quark mass, 
the left--handed $Z^0$ coupling shift, due to the gauge boson mixing 
and the mixing of the SM $b$ quark with the unique $b'$ state, satisfies 
$\delta g_{Z^0}^{b_L}/g_{Z^0}^{b_L} \vert_{{\rm TOTAL}} \lesssim - 1.5\%$.
\\ Such a shift at $\delta g_{Z^0}^{b_L}/g_{Z^0}^{b_L} \vert_{{\rm TOTAL}} \sim - 1.5\%$ 
with respect to SM is possibly acceptable in the sense that it leads to
theoretical predictions in a good agreement with measurements of $R_b$ and
$A^b_{FB}$, if at the same time $\delta g_{Z^0}^{b_R^c}/g_{Z^0}^{b_R^c} \vert_{{\rm TOTAL}}$ is positive
and large (see quantitative discussion at the beginning of Section \ref{Third}).
However, $\delta g_{Z^0}^{b_R^c}/g_{Z^0}^{b_R^c} \vert_{{\rm TOTAL}}$ is either negative or
too small so that this minimal model is excluded if one searches to address the
$A^b_{FB}$ anomaly. 
\\ Furthermore, the limit on the theoretical shift 
$\delta g_{Z^0}^{b_L}/g_{Z^0}^{b_L} \vert_{{\rm TOTAL}} \lesssim - 1.5\%$ obtained in \cite{LRarg} 
does not hold for the other models we are considering in the present paper, due to the following 
main different features. 
First, in the several models studied in this paper, we consider representations which are different from Eq.(\ref{Min:Q1LtcR})-(\ref{Min:Q2LbcR}) 
and give rise to various sign configurations for the individual $Z^0$ coupling shifts. For instance, in our Models I and II, one finds
$\delta g_{Z^0}^{b_L} / g_{Z^0}^{b_L} \vert_{{\rm boson}} <0$ and $\delta g_{Z^0}^{b_L} / g_{Z^0}^{b_L} \vert_{{\rm fermion}} >0$ (in contrast
with the minimal model where both are negative) so that a compensation occurs tending to decrease $\vert \delta g_{Z^0}^{b_L}/g_{Z^0}^{b_L} \vert_{{\rm TOTAL}} \vert$.
Moreover, e.g. in Model II, several $b'$,$t'$ states [with electric charges $-1/3$,$2/3$ and boundary conditions $(-+)$] 
are introduced in contrast with Eq.(\ref{Min:Q1LtcR})-(\ref{Min:Q2LbcR}).
Secondly, in this paper we do not assume $\tilde g=g$ (fixing $g_{Z'}$) and we consider a specific mechanism (mixing $Q_{1L}$--$Q_{2L}$), 
which allows more freedom on the parameter space than for the minimal model.
\\ \\ 
{\bf $O(3)$ inspired:}
The model suggested in \cite{AFB-Rold} is characterized by the symmetry ${\rm O(3) \equiv SU(2)_V \times P_{LR}}$ [${\rm P_{LR}}$ being a left--right parity] 
and the associated permitted representations are, 
\begin{eqnarray}
\{Q_{L}\} \ \equiv \ 
({\bf 2},{\bf 2})_{2/3} \ = \
\left(\begin{array}{cc} q^{\prime}_{{\rm (5/3)} L} & t_{L} \\ t^{\prime}_L & b_{L} \end{array}\right) 
\ \ \ 
\{t^c_R\} \ \equiv \ 
({\bf 1},{\bf 1})_{2/3} \ = \
\left(\begin{array}{c} t^{c}_R \end{array}\right)
\label{rep:O3}
\end{eqnarray}
or $\{ t^c_R \} \equiv ({\bf 1},{\bf 3})_{2/3} \oplus ({\bf 3},{\bf 1})_{2/3}$ [or even $\{ t^c_R \} \equiv ({\bf 3},{\bf 3})_{2/3}$]. 
The motivation for imposing this symmetry was that it protects
the $Z^0$ vertex by insuring 
$\delta g_{Z^0}^{b_L} / g_{Z^0}^{b_L} \vert_{{\rm fermion}} = 0$ and 
$\delta g_{Z^0}^{b_L} / g_{Z^0}^{b_L} \vert_{{\rm boson}} = 0$. 
However, for such a shift $\delta g_{Z^0}^{b_L} / g_{Z^0}^{b_L} \vert_{{\rm TOTAL}} = 0$, 
the global fit of $R_b$ and
$A^b_{FB}$ cannot be significantly improved compared to the SM case, whatever is the value of $\delta g_{Z^0}^{b_R^c}/g_{Z^0}^{b_R^c} \vert_{{\rm TOTAL}}$ 
\cite{RSAFB}. The story does not finish here as there are a few sources of small contributions to $\delta g_{Z^0}^{b_L} / g_{Z^0}^{b_L} \vert_{{\rm TOTAL}}$.
The first source is the breaking of ${\rm SU(2)_R}$ through boundary conditions which in turn violate the custodial symmetry subgroup protecting the SM
$Z^0 \bar b_L b_L$ coupling. Numerically, this gives e.g. $\delta g_{Z^0}^{b_L} / g_{Z^0}^{b_L} \vert_{{\rm boson}} \simeq 0.14 \%$ for $M_{KK}=3$ TeV. 
$\delta g_{Z^0}^{b_L} / g_{Z^0}^{b_L} \vert_{{\rm boson}}$ can also receive a contribution from the breaking of ${\rm P_{LR}}$ due to different
coupling constants associated to ${\rm SU(2)_R}$ and ${\rm SU(2)_L}$: $\tilde g \neq g$.
\\ As we have demonstrated in the beginning of Section \ref{Third}, if one wants, at the same time, to reproduce the correct $b,t$ masses and to
improve the fit on $R_b$,$A^b_{FB}$, the introduced mechanism mixing $Q_{1L}$ with $Q_{2L}$ must be invoked. This is true also in this ${\rm O(3)}$ symmetry context. 
In order to generate a top quark mass, one takes $\{Q_{1L}\}=\{Q_{L}\}$ (see Eq.(\ref{rep:O3}) and Eq.(\ref{5Daction})). $\{Q_{2L}\}$ and $\{b^c_R\}$ 
embeddings are then chosen in order to
arrange the fit of $R_b$ and $A^b_{FB}$. In general, this choice leads to a breaking of the custodial symmetry subgroup, which protects the 
$Z^0 \bar b_L b_L$ coupling, via the bottom Yukawa coupling sector
\cite{AFB-Rold} (it happens to be the case for the models considered in this
paper). This breaking gives a contribution to $\delta g_{Z^0}^{b_L} / g_{Z^0}^{b_L} \vert_{{\rm TOTAL}}$ and it is naturally weak
since the ratio $m_b/m_t$ has to be small (this is visible through the $\cos\theta$ suppression in the element $1,1$ of mass matrix (\ref{MbTOTAL})). 
\\ A slightly different motivation thus arises for the ${\rm O(3)}$ symmetry: the smallness of $\delta g_{Z^0}^{b_L} / g_{Z^0}^{b_L} \vert_{{\rm TOTAL}}$ 
($\sim -1\%$ required) could be explained by the weak breaking effects of the custodial symmetry subgroup described above [in analogy with the
't Hooft criteria of naturality]. Nevertheless, we have found quantitatively, by taking into account all the above breaking effects contributing to the necessary
$\delta g_{Z^0}^{b_L} / g_{Z^0}^{b_L} \vert_{{\rm TOTAL}}$, that there exist no complete model of this type (\ref{rep:O3}) respecting the ${\rm O(3)}$ symmetry 
(slightly violated in the bottom sector) which is able to simultaneously
reproduce $m_{b,t}$ and solve the whole $A^b_{FB}$ anomaly [imposing $10 \%  $ on the fit].
The reason is that the produced value of $|\delta g_{Z^0}^{b_L} / g_{Z^0}^{b_L} \vert_{{\rm TOTAL}}|$ 
appears systematically to be too small. This is mainly explained by the following
considerations. For certain $\{Q_{2L}\}$ representations, 
$\delta g_{Z^0}^{b_{2L}} / g_{Z^0}^{b_{2L}} \vert_{{\rm boson}} < 0 $ is
generally realizable in the bottom sector [${\rm O(3)}$ breaking]. Nevertheless, 
the whole induced $|\delta g_{Z^0}^{b_{L}} / g_{Z^0}^{b_{L}} \vert_{{\rm boson}}|$
({\it c.f.} element $1,1$ of matrix (\ref{NCinter}) with $\delta g_{Z^0}^{b_{1L}} \vert_{{\rm boson}} \simeq 0$ 
due to ${\rm O(3)}$) is too much suppressed by the $\cos^2\theta$ factor [too weak ${\rm O(3)}$ violation]
imposed from the acceptably small $m_b/m_t$ amount. To be noted that, staying
with multiplets smaller than ${\bf 4}$, the $|\delta g_{Z^0}^{b_{L}} / g_{Z^0}^{b_{L}} \vert_{{\rm fermion}}|$
contribution is also not sufficiently large due to the $\{b^c_R\}$ multiplet structure [to
which possibly belong the $b^{c\prime(1)}_R$ having a direct mixing mass term with $b^{(0)}_{2L}$]
that is needed to generate a satisfactory $\delta g_{Z^0}^{b_R^c}/g_{Z^0}^{b_R^c} \vert_{{\rm TOTAL}}$.
\\ \\ 
{\bf Model IV:}
In order to show that it is possible to construct a model, with some representation larger than ${\bf 3}$, allowing to fulfill both conditions of EW fit improvement and mass
reproduction, we simply present an example of model that we find to be of this kind. 
It consists of a multiplet extension just from the top quark sector of Model III, introducing no new $b'$ or $t'$ state:
\begin{eqnarray}
\{Q_{1L}\} \ \equiv \ 
({\bf 2},{\bf 3})_{7/6} \ = \
\left(\begin{array}{ccc} q^{\prime}_{{\rm (8/3)} L} & q^{\prime}_{{\rm (5/3)} L} & t_{1L} \\ q^{\prime\prime}_{{\rm (5/3)} L} & t^{\prime}_L & b_{1L} \end{array}\right) 
\ \ \ 
\{t^c_R\} \ \equiv \ 
({\bf 1},{\bf 4})_{7/6} \ = \
\left(\begin{array}{cccc} q^{c\prime}_{{\rm (8/3)} R} & q^{c\prime}_{{\rm (5/3)} R} & t^c_R & b^{c\prime}_R \end{array}\right)
\label{repIV:Q1LtcR}
\end{eqnarray}
We find, once more, $\delta g_{Z^0}^{b_R^c} / g_{Z^0}^{b_R^c} \vert_{{\rm boson}}>0$, $\delta g_{Z^0}^{b_R^c} / g_{Z^0}^{b_R^c} \vert_{{\rm fermion}}>0$ 
and $\delta g_{Z^0}^{b_{R}^c}/g_{Z^0}^{b_{R}^c} \vert_{{\rm TOTAL}} > 0$, while
$\delta g_{Z^0}^{b_L} / g_{Z^0}^{b_L} \vert_{{\rm boson}} >0$, $\delta g_{Z^0}^{b_L} / g_{Z^0}^{b_L} \vert_{{\rm fermion}} <0$ 
and $\delta g_{Z^0}^{b_{L}}/g_{Z^0}^{b_{L}} \vert_{{\rm TOTAL}} < 0$.
\\ The quantitative results and plots that we obtain for the present model are comparable to those obtained for Model III. The small differences originate
from two sources: 
(a) different Clebsch--Gordan coefficients induce new mass matrix elements and in turn deviations on $m_{b,t}$ as well as on $\delta g_{Z^0}^b / g_{Z^0}^b \vert_{{\rm fermion}}$
(b) a different quantum number for $b_{1L}$, namely its isospin $I_{3R}^{b_{1L}}$, is modified so that 
$\delta g_{Z^0}^{b_L} / g_{Z^0}^{b_L} \vert_{{\rm boson}}$ is affected, via $Q_{Z'}^{b_{1L}}$.

\section{Conclusion}

In the RS framework with a bulk custodial symmetry, we have elaborated the set
of realistic models, defined by fermion representations smaller than ${\bf 4}$ and 
based on the necessary mechanism introducing two left--handed doublets $Q_{iL}$, which can explain precision EW data in the heavy quark sector. 
Indeed, these models allow to solve the $A^b_{FB}$ anomaly at the $Z^0$ pole and outside the resonance domain - thanks to KK fermion and boson mixing effects -
while keeping the predictions for $R_b$ and $m_{b,t}$ in good agreement with experimental values. 
More precisely, the global fit of $R_b$ and $A^b_{FB}$ [at the various center of mass energies] is significantly improved with respect to the pure SM case,
for $M_{KK}=3,4,5$ TeV (higher energy scales were not considered to not worsen the remaining little hierarchy problem).
In the heavy Higgs regime, $m_h \gtrsim 500$ GeV, this improvement is not systematically significant.
\\ Moreover, we have shown that the obtained models can simultaneously lead also to an important improvement (compared to SM) of the fit on EW observables in the light fermion and gauge boson sector. 
This was done by considering a fix set of values, for the fundamental parameters entering both the $A^b_{FB}$ and oblique parameter analyses: 
$M_{KK}$ and $g_{Z'}$ (as well as the sign determining $\sin^2 \theta'$), being common to both analyses.
\\ As a matter of fact, concerning the EW fit, the theoretical estimations for $S_{\rm RS}$,$T_{\rm RS}$ can reach domains of higher degree of agreement 
(relatively to SM) in the case of the obtained models, 
for $M_{KK}=3-5$ TeV, any allowed $g_{Z'}$ value and $m_h \geq 115$ GeV.
In particular, if $m_h=500$ GeV, while the EW fit is at a dramatic degree of agreement of $2.5 \ 10^{-9}$ in the SM, it can still be acceptable 
in the RS scenario [e.g. at $25.3 \%$ for $M_{KK}=4$ TeV, $g_{Z'}=1.25$] opening up the possibility of an heavy Higgs boson regime.
\\ Besides, the best EW fit can correspond to an Higgs mass such that $m_h \geq 115$ GeV, allowing then to respect the LEP2 lower bound, 
which is impossible within the SM framework. 

\vspace{0.8cm}

Inside the relevant domains of parameter space (where EW fits are improved and $m_{b,t}$ are reproducable), custodians [states like $b'$, $q^{c\prime}_{{\rm (5/3)}}$,\dots] 
can reach quite low masses. As discussed, such particles could be as low as $\sim 1200$ GeV which, in view of the LHC physics, is considerably smaller than the typical favored 
value for the first KK gauge boson mass: $M_{KK} \sim 4$ TeV. Therefore, within the models obtained here, the (single) custodian production \cite{LHCfermion,Houches} might lead to 
stronger effects and more visible signatures at LHC, relatively to the production of KK excitations of gauge bosons \cite{RSAFBLHC,LHNC,HRizzoBIS,Houches,LHCboson}. 
\\ This conclusion illustrates the fact that precision EW data may continue to
benefit from a certain emphasis in the future. Indeed, the constraints on a given parameter space coming from 
EW data serve as a guide for testing new physics at the LHC. 
Moreover, if some physics beyond the SM is discovered at an high--energy collider, one would have to study the new physics effects on EW fits (possibly including results from 
the $Z^0$ boson factory project Giga--Z); 
a global interpretation [e.g. through $b'$ effects] of both the high--energy data and the precision EW data would then constitute a solid confirmation of the theoretical picture.
\\
\\
\noindent \textbf{Acknowledgments:} 
The authors are grateful to K.~Agashe, M.~Boonekamp, R.~Contino, A.~Djouadi, A.~Goudelis and F.~Richard 
for useful discussions, in particular at Les Houches Workshop \cite{Houches}.
We also thank M.~Calvet for her contribution to the manuscript.

\end{document}